\documentclass[journal]{IEEEtran}
\usepackage{graphicx,cite,epsfig,amssymb,amsmath,subfigure,url,stfloats,latexsym}
\usepackage{bm}
\usepackage{algorithm}
\usepackage{algorithmic}
\usepackage{cases}
\usepackage{color}
\usepackage{epstopdf}

\newtheorem{theorem}{Theorem}

\newtheorem{lemma}{Lemma}

\newtheorem{corollary}{Corollary}

\newtheorem{proposition}{Proposition}
\newtheorem{definition}{Definition}

\usepackage[left=0.6in,right=0.6in,bottom=1in,top=0.7in]{geometry}
\begin{document}

\title{SCMA Codebook Design Based on Uniquely Decomposable Constellation Groups}

\author{Xuewan~Zhang,
        Dalong~Zhang,
        Liuqing~Yang,~\IEEEmembership{Fellow,~IEEE,}
        Gangtao~Han,
        Hsiao-Hwa~Chen,~\IEEEmembership{Fellow,~IEEE,}
        Di~Zhang,~\IEEEmembership{Senior Member,~IEEE}

      \thanks{Manuscript received April 1, 2020; revised July 13, 2020 and December 1, 2020; accepted February 17, 2021.}
      \thanks{This study was supported by the National Science Foundation of China (NSFC) under grant: 62001423, the Henan Provincial Key Research, Development and Promotion Project under grant: 212102210175, 202102210123; the Henan Provincial Key Scientific Research Project for College and University under grant: 21A510011.}
      \thanks{Xuewan Zhang, Dalong Zhang, Gangtao Han and Di Zhang are with the School of Information Engineering, Zhengzhou University, Zhengzhou 450001, China (email: iexwzhang@gs.zzu.edu.cn, iedlzhang@zzu.edu.cn, iegtHan@zzu.edu.cn, dr.di.zhang@ieee.org).}
      \thanks{Liuqing Yang is with the Department of Electrical and Computer Engineering, University of Minnesota, Minneapolis, MN 55455, USA (e-mail: qingqing@umn.edu).}
      \thanks{Hsiao-Hwa Chen is with the Department of Engineering Science, National Cheng Kung University, Tainan 70101, Taiwan (e-mail: hshwchen@mail.ncku.edu.tw).}
      \thanks{Corresponding author: Di Zhang (email:dr.di.zhang@ieee.org).}
}

\markboth{IEEE Transactions on Wireless Communications, Vol. XX, No. YY, Month 2021}
{Zhang, \emph{et al.}: SCMA Codebook Design \ldots}

\IEEEpubid{\begin{tabular}[t]{@{}l@{}}\centerline{}\\\centerline{}\\\centerline{}\\
		\centerline{Copyright (c) 2021 IEEE. This is a preprint version. Personal use of this material is permitted.}\\ \centerline{However, permission to use this material for any other purposes must be obtained from the IEEE by sending a request to pubs-permissions@ieee.org.}\end{tabular}}

\maketitle



\begin{abstract}
Sparse code multiple access (SCMA), which helps improve spectrum efficiency (SE) and enhance connectivity, has been proposed as a non-orthogonal multiple access (NOMA) scheme for 5G systems. In SCMA, codebook design determines system overload ratio and detection performance at a receiver. In this paper, an SCMA codebook design approach is proposed based on uniquely decomposable constellation group (UDCG). We show that there are $N+1$ ($N\ge 1$) constellations in the proposed UDCG, each of which has $M (M\ge 2)$ constellation points. These constellations are allocated to users sharing the same resource. Combining the constellations allocated on multiple resources of each user, we can obtain UDCG-based codebook sets. Bit error ratio (BER) performance will be discussed in terms of coding gain maximization with superimposed constellations and UDCG-based codebooks. Simulation results demonstrate that the superimposed constellation of each resource has large minimum Euclidean distance (MED) and meets uniquely decodable constraint. Thus, BER performance of the proposed codebook design approach outperforms that of the existing codebook design schemes in both uncoded and coded SCMA systems, especially for large-size codebooks.
\end{abstract}


\vspace{0.35in}
\IEEEpeerreviewmaketitle

\section{Introduction}
The growing demands for high-speed, low-latency, massive connectivity services have posed a big challenge to today's wireless communications \cite{wang2018device,yang2018internet,fan2019angle}. The fifth generation (5G) and beyond communications need to support a very high spectrum efficiency (SE) per area \cite{HongApp}. Due to limited spectrum resources, existing orthogonal frequency division multiple access (OFDMA) technique cannot meet these requirements. Non-orthogonal multiple access (NOMA) \cite{liu2017enhancing,mao2018rate}, on the other hand, promises to transmit data for multiple users with the same resource block, which has attracted an increasing attention from both academia and industry \cite{dzhang19,hu2019application,zhang2018power, JTang,SVC2020Zhang}.

Typically, NOMA can be divided into power domain NOMA \cite{JTang} and code domain NOMA \cite{mao2018rate}. In this work, we focus on one of code domain NOMA, i.e., sparse code multiple access (SCMA) \cite{nikopour2013sparse}. SCMA transmits data of multiple users on the same frequency resources, and each user may access multiple frequency resources for its data transmission, which improves systems' SE \cite{MaxYu}. In SCMA systems, each user will be pre-allocated a unique codebook consisting of complex multi-dimensional codewords. The bit steam of each user is mapped directly into the resources with these codewords. Due to the sparsity of SCMA, multi-user detector can use a low-complexity message passing algorithm (MPA) \cite{zhang2018low,ameur2019performance,du2016fast,dai2019iterative} to obtain a near optimal bit error ratio (BER) performance. However, the number of codebooks is relatively small, making it very hard to achieve a high overload ratio. In addition, users are distinguished by their pre-allocated codebooks in SCMA systems. Therefore, codebook design is one of the most critical issues in SCMA systems.

Several works have been reported on SCMA codebook design. For instance, the original codebook (OCB) design scheme was proposed first in \cite{taherzadeh2014scma}, where a multi-dimensional complex mother constellation was generated by Cartesian Product of two quadrature amplitude modulation (QAM) constellations. With the mother constellation, a unique codebook set of users can be obtained using different permutation and complex conjugate operations. However, the permutation set used in OCB scheme was not optimal, leading to a sub-optimal power diversity gain among different codebooks. To obtain an optimal SCMA codebook, other codebook design schemes based on constellation operations were suggested in \cite{zhou2017scma,dong2018efficient,cai2016multi,yan2017dimension,gao2017low,alam2017performance}. The authors in \cite{zhou2017scma} proposed a constellation rotation-based codebook (CRCB) design scheme. The authors of \cite{dong2018efficient} found that a larger minimum Euclidean distance (MED) can be achieved among codewords to optimize the rotation angle with the channel parameters. Different from \cite{zhou2017scma} and \cite{dong2018efficient}, interleaving was used to enlarge the MED among codewords of a codebook in \cite{cai2016multi}. In addition to phase rotation, coordinate interleaving and permutation operations, the authors in \cite{yan2017dimension} presented a design with Trellis coded modulation (TCM). Low density signature (LDS) with constellation rotation-based codebook (LCRCB) can achieve a better BER performance, as confirmed in \cite{gao2017low,alam2017performance}. Although the BER performance of the aforementioned schemes can be improved with constellation operations, the constellation related performance has been rarely considered.

There are yet other works combining constellation to design SCMA codebooks \cite{metkarunchit2017scma,liu2018optimized,yu2018design,mheich2018design}. The work in \cite{metkarunchit2017scma} proposed a design scheme to create multi-dimensional mother constellation of SCMA codebooks with circular QAM. Based on circular QAM, an optimized SCMA codebook design scheme with QAM segmentation was presented in \cite{liu2018optimized}. Similar to \cite{liu2018optimized}, the authors in \cite{yu2018design} proposed an optimized codebook design scheme based on star QAM. The difference between the works in \cite{liu2018optimized} and \cite{yu2018design} lies on the SCMA codebook design criteria. In \cite{liu2018optimized}, maximized the MED of circular QAM was used to design the SCMA codebooks, while error probability of superimposed constellation points was considered in \cite{yu2018design}. Besides circular QAM, a golden angle modulation based codebook (GAMCB) design was presented in \cite{mheich2018design}, where a low peak to average power ratio (PAPR) of a codebook was used as the design criteria. However, the low-PAPR codebook design may reduce power diversity gain between codewords. In addition, a small codebook size was assumed in most studies, such as four points, which may not guarantee a satisfactory BER performance while codebook size grows. Moreover, most of these approaches cannot guarantee a maximum coding gain based on the designed codebooks.

To tackle the aforementioned problems, a class of lattice-based codebook (LCB) design was proposed \cite{yu2016novel,bao2016spherical,yan2016top,zhang2019efficient}. Studies in \cite{yu2016novel} and \cite{bao2016spherical} proposed two novel SCMA design schemes with a maximized shaping gain of the constellation. Different from \cite{yu2016novel} and \cite{bao2016spherical}, the works in \cite{yan2016top} and \cite{zhang2019efficient} considered the coding gain and shaping gain jointly. In \cite{yan2016top}, the constellation generated by an optimal two-dimensional lattice structure was decomposed into three constellations, which are allocated to three users on the same resource to realize a 150\% SCMA overload ratio. The authors in \cite{zhang2019efficient} designed the codebooks in a multi-dimensional space with an optimal multi-dimensional lattice structure. Although LCB design can improve the BER performance, the MED of superimposed constellations is zero, which means that the superimposed constellation of each resource is not decodable and some codewords in LCB scheme may be distinguished only by a single resource, not multiple resources. For this reason, multi-user detector may not be use the desired codewords, leading to an unsatisfactory BER performance.

\subsection{Motivation and Contributions}
As mentioned earlier, SCMA codebook design has been investigated from different perspectives. However, there are still some problems to be solved, which motivate us to develop this paper. In order to make the superimposed constellations of each resource decodable and to facilitate multi-user detection of multi-dimensional codewords, an SCMA codebook deign approach is proposed based on uniquely decomposable group (UDCG) in this study. Generally speaking, several small constellations can be superimposed to form a large constellation, namely superimposed constellation. If any point in the superimposed constellation uniquely corresponds to the points in these small constellations, we treat the superimposed constellation as a uniquely decodable constellation (UDC). Essentially, the UDCG defines the UDC relationship between a constellation group and its superimposed constellation. Although an exhaustive method based on Lagrange four-square theorem to select points satisfying UDCG constraints was introduced in \cite{xiong2012energy}, this method is not applicable for multiple constellations with a large number of points. Based on the results in \cite{xiong2012energy}, \cite{han2017uniquely} divided the QAM constellation into two phase shift keying (PSK) constellations to sever two users. Moreover, the authors in \cite{li2018noncoherent} proposed a uniquely decomposable constellation pair (UDCP) based on PSK constellations, and gave a simple proof for the UDCP. However, the number of constellations in the proposed UDCG is small and each constellation has only a few constellation points, which is not suitable for SCMA codebook design.

To overcome these problems, we propose a new UDCG in this paper. It contains several constellations with a large number of constellation points, based on which we introduce an SCMA codebook design approach. In the UDCG-based codebook design, these constellations contained in the UDCG are allocated to users to establish a codebook generation matrix, which can be used to generate the codebooks according to a pre-designed factor graph matrix \cite{zhou2017scma,zhang2019efficient}. Thereafter, we will discuss the implementation issue for UDCG-based codebooks with a desired BER constraint as an optimization problem. We give its optimal solution with two methods, i.e., maximized coding gain with superimposed constellation, and maximized coding gain with codebooks. Due to the uniquely decodable property of the UDCG-based constellation, the MPA multi-user detector can make a full use of multi-dimensional codewords for joint detection to improve BER performance. The main contributions of this work can be summarized as follows.
\begin{enumerate}
\item We propose a constellation group meeting the UDCG constraint, which contains $N+1\ (N \geqslant 1)$ constellations, with each having $M$ ($M \geqslant 2$) constellation points to suit for the SCMA codebook.
\item We propose a low-complexity UDCG-based SCMA codebook design scheme, which allocates the constellations according to a pre-designed rule as suggested in \cite{zhou2017scma}. The proposed SCMA codebooks guarantee that the superimposed constellation of each resource meets the UDC constrains.
\item We formulate a BER performance optimization problem with the UDCG-based codebook design scheme. The problem is solved sequentially by maximizing the coding gain of superimposed constellation and the codebook. The large MED among codewords and among superimposed constellation points are guaranteed as well.
\item Over Gaussian and Rayleigh fading (RF) channels, BER performance of the proposed UDCG-based codebooks is verified in both uncoded and coded SCMA systems. Simulation results show that the proposed codebooks have good BER performance, especially for large-size codebooks.
\end{enumerate}

The remainder of this paper are outlined as follows. Section II introduces the system model and its problem formulation. The SCMA codebook design based on UDCG is illustrated in Section III, where the basics of the UDCG is given first, followed by the designs of the proposed UDCG and UDCG-based codebook. In Section IV, an implementation issue about UDCG-based codebooks is discussed in terms of single and multiple resources, respectively. Section V is the simulation results, followed by the conclusions in Section VI.

\vspace{0.125in}
\section{System model and problem formulation}
\subsection{SCMA System Model}
\begin{figure}[t]
  \centering
  \includegraphics[width=0.48\textwidth]{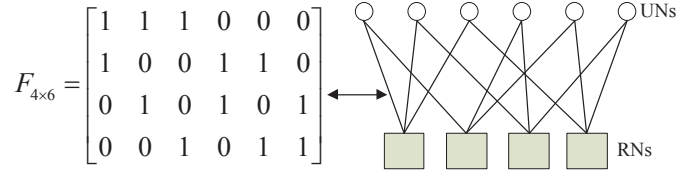}
  \caption{The mapping relationship between UNs and RNs characterized by $\bm F_{4 \times 6}$.}
  \label{UNvsRN}
\end{figure}

Let us consider a downlink SCMA system with $J$ user nodes (UNs) and $K$ ($J>K$) resource nodes (RNs), whose overload ratio is defined as $\lambda  = J/K$. The structure of $J$ UNs sharing $K$ RNs can be represented by a factor graph matrix $\bm F = \left[ {{\bm f_1}, \cdots ,{\bm f_J}} \right]$. Each column of $\bm F$ has $B \left( {B < K} \right)$ non-zero elements (1), which indicates that the signal of each UN is carried by $B$ RNs. The number $d_f$ of elements 1 in each row of $\bm F$ denotes how many UNs are sharing the same RN, and the value of ${d_f}$ is given as \cite{nikopour2013sparse}
\begin{equation} \label{F：parameters}
{d_f}=\Big( \begin{array}{c}
{K - 1} \\
{B - 1} \\
\end{array}
\Big)
= {{JB} \over K}.
\end{equation}
Here, an example to explain the mapping relationship between UNs and RNs with $K=4$, $J=6$ and $B=2$ is shown in Fig. \ref{UNvsRN}.

Assume that $\mathcal {\bm X}_j \in {{\Bbb C}^{K \times M}}$ denotes the $j$-th UN's codebook, and the $m_j$-th $\left( {{m_j} = 1,2, \cdots M} \right)$ column ${\bm x_{m_j}} = {\left[ {x_{1,m_j}, \cdots ,x_{K,m_j}} \right]^T}$ of $\mathcal {\bm X}_j$, which is a codeword, consists of $B$ non-zero complex elements and $K-B$ zero elements. The positions of these elements in $\{ {{\bm x_{m_j}}} \}_{m_j = 1}^M$ correspond to $\bm f_j$. In the downlink, a codeword ${\bm x_{m_j}}$ of $j$-th UN's $\log2(M)$ bits is obtained via its codebook $\mathcal {\bm X}_j$, and after the synchronous layer multiplexing, the received signal ${\bm y}$ will be \cite{mheich2018design}

\begin{align}
{\bm y}=\sum_{j=1}^J{{\rm diag}\left(\bm h \right)}{\bm x}_{m_j}+{\bm z},
\label{y1}
\end{align}
where $\bm h = {\left[ {{h_1}, \cdots ,{h_K}} \right]^T}$ with $h_k \sim \mathcal{CN}(0,1)$ is the channel vector, and ${\bm z}=[z_1,\cdots,z_K]^T$ denotes the additive white Gaussian noise (AWGN) channel that obeys the complex Gaussian distribution $\mathcal{CN}(0,\sigma^2 {\bm I})$.

Moreover, if we assume all $\{h_k \}_{k = 1}^K$ independent, the received signal for the $k$-th RN can be expressed as
\begin{equation}
{y_k} = \sum\limits_{j \in {\xi _k}} {{h_k}x_{k,m_j}}+ {z_k},
\label{y2}
\end{equation}
where $\xi _k$ represents the non-zero positions in $k$-th row of the factor matrix $\bm F$.

\subsection{Problem Formulation}
Let $\mathcal {\bm X}= {\mathcal {\bm X}}_1 \times  \cdots  \times {\mathcal {\bm X}_J}$ denote the combination constellation of all UNs' consebooks, $\bm H={\rm diag}\left(\bm h \right)$ and the transmitted vector ${\bm x^{(i_1)}} = {\left[ {x_1^{(i_1)}, \cdots ,x_K^{(i_1)}} \right]^T}$ belong to $\mathcal {\bm X}$. Giving any vector $\bm x^{(i_2)} \in {\mathcal {\bm X}}, (i_2 \ne i_1)$, the pairwise error probability (PEP) between $\bm x^{(i_1)}$ and $\bm x^{(i_2)}$ will be
\begin{equation}
\begin{aligned}
P_e \left( {{\bm x^{(i_1)}} \to {\bm x^{(i_2)}}\left| \bm H \right.} \right) = Q\left( {\sqrt {\frac{\left\| {\bm H \left( {{\bm x^{(i_1)}} - {\bm x^{(i_2)}}} \right)} \right\|_2^2}
{2{\sigma ^2}}} } \right),
\end{aligned}
\label{PEP}
\end{equation}
where $Q\left(  \cdot  \right)$ denotes the Q-function. With (\ref{PEP}) in hand, the error rate of transmitted vector $\bm x^{(i_1)}$ can be given as
\begin{equation}
\begin{aligned}
& P_e \left( {{\bm x^{(i_1)}}\left| \bm H \right.} \right) = \sum\limits_{{i_2} = 1,{i_2} \ne {i_1}}^{{M^J}} {{P_e}\left( {{\bm x^{\left( i_1 \right)}} \to {\bm x^{\left( i_2 \right)}}\left| \bm H \right.} \right)} \\
&= \sum\limits_{i_2=1,i_2 \ne i_1}^{M^J} {Q\left( {\sqrt {\frac{{\sum\limits_{k = 1}^K {{{\left| {h_k} \right|}^2}{{\left| {x_k^{(i_1)} - x_k^{(i_2)}} \right|}^2}} }}
{2{\sigma ^2}}} } \right)}.
\end{aligned}
\label{ERR}
\end{equation}
Let $\Xi_k=\sum\limits_{j \in {\xi _k}} {x_{k,{m_j}}} $ denote the superimposed constellation at the $k$-th RN, and $d_{\min}$ denotes the MED of $\Xi_k, \forall k=1,2, \cdots, K$. We then have $x_k^{(i_1)}, x_k^{(i_2)} \in \Xi_k$ and $\left| {x_k^{(i_1)} - x_k^{(i_2)}} \right| \geqslant d_{\min}, \forall i_1,i_2, i_2 \ne i_1$. In this case, the upper bound of $P_e \left( {{\bm x^{(i_1)}}\left| \bm H \right.} \right)$ in (\ref{ERR}) can be given as \cite{yu2018design}
\begin{equation}
\begin{aligned}
P_e \left( {{\bm x^{({i_1})}}\left| \bm H \right.} \right) \leqslant \left( {{M^J} - 1} \right)Q\left( {\sqrt {\frac{{\sum\limits_{k = 1}^K {{{\left| {h_k} \right|}^2}} }}
{{2{\sigma ^2}}}} {d_{\min }}} \right).
\end{aligned}
\label{UPB1}
\end{equation}
Furthermore, by following average inequality $\frac{{\sum\limits_{k = 1}^K {{{\left| {x_k^{({i_1})} - x_k^{({i_2})}} \right|}^2}} }}{K} \geqslant \root K \of {\prod\limits_{k = 1}^K {{{\left| {x_k^{({i_1})} - x_k^{({i_2})}} \right|}^2}} }$, the upper bound of (\ref{ERR}) in an AWGN channel is
\begin{equation}
\begin{aligned}
P_e \left( {\bm x^{({i_1})}} \right) \leqslant \left( {{M^J} - 1} \right)Q\left( {\sqrt {\frac{K}
{2{\sigma ^2}}} d_{\min }} \right).
\end{aligned}
\label{UPB2}
\end{equation}
From (\ref{UPB1}) and (\ref{UPB2}), we can know that the success detection probability of SCMA is determined by $d_{\min}$ for both AWGN and RF channels. When $d_{\min}=0$, (\ref{UPB2}) has the maximum value. This indicates that some points in $\Xi_k$ are repeated, which makes it difficult to distinguish $x_{k,m_j}$ from those that belong to a repeated point. In this paper, we consider that the constellation $\Xi_k$ with $d_{\min}=0$ is not unique decodable; otherwise we call the constellation as an UDC when $d_{min}>0$, and the definition of UDC is thus given as
\begin{definition}
Giving arbitrary two point $x_k^{(i_1)}$ and $x_k^{(i_2)}$ of constellation $\Xi_k$, the constellation $\Xi_k$ is an UDC if $x_k^{(i_1)} \ne x_k^{(i_2)}$ with $i_1 \ne i_2$.
\label{def0}
\end{definition}

According to (\ref{UPB1}), (\ref{UPB2}) and Definition \ref{def0}, increasing $d_{\min}$ will bring in a large success probability of SCMA detector. However, most of the existing codebook designs only considered the MED among codewords, and ignored the MED among superimposed constellation points of each RN. The undecodable superimposed constellation will restrict the detection performance. To further improve the BER performance, our goal of this work is to design a set of SCMA codebooks whose superimposed constellation has a large $d_{\min}$.

\vspace{0.125in}
\section{SCMA codebook design based on UDCG}
In this section, we will elaborate the proposed UDCG-based SCMA codebook design scheme
to solve the aforementioned problems. With the proposed UDCG, we formulate the rules of constellation allocation for UNs on the same RN to obtain the UDCG-based SCMA codebooks.

\subsection{Basics of Uniquely Decomposable Constellation Group}
In general, we assume that the $n$-th constellation ${{\mathcal S}_n}, \forall n \in \{ 0,1, \ldots ,N\}$ involved in this section is regarded as a vector. The UDCG concept was initially proposed in \cite{xiong2012energy}, which can be defined as
\vspace{0.1in}
\begin{definition}
Let ${s_n},{\tilde s_n} \in {{\mathcal S}_n}$ be two constellation points of the $n$-th constellation ${{\mathcal S}_n},\forall n \in \{ 0,1, \ldots ,N\}$. If and only if ${s_n} = {\tilde s_n}$ satisfy $\sum _{n = 0}^N{s_n} = \sum _{n = 0}^N{\tilde s_n}$, we say that $\{ {{\mathcal S}_n}\} _{n = 0}^N$ form an UDCG, which is denoted by $ \uplus _{n = 0}^N{{\mathcal S}_n} = {{\mathcal S}_0} \uplus {{\mathcal S}_1} \uplus  \ldots {{\mathcal S}_N}$.
\label{def1}
\end{definition}
\vspace{0.1in}

Here, $ \uplus _{n = 0}^N{{\mathcal S}_n}$ denotes the superimposed constellation and ${{\mathcal S}_n},\forall n \in \{ 0,1, \ldots ,N\}$ is a sub-constellation of UDCG. According to Definition \ref{def0}, we know that Definition \ref{def1} explains a unique correspondence between the superimposed constellation and the sub-constellations. In other words, the superimposed constellation of $N+1$ sub-constellations meets the UDC constraint. Based on Definition \ref{def1}, the authors in \cite{li2018noncoherent} proposed a constellation pair meets the UDCP constraint. When $N=1$, the UDCP definition can be expressed as follows.
\vspace{0.1in}
\begin{definition}
For any ${\alpha _0} > 0$ and ${\alpha _1} > 0$, if and only if ${s_{0}} = {\tilde s_{0}}$ and ${s_{1}} = {\tilde s_{1}}$ such that ${\alpha _0}{s_{0}} + {\alpha _1}{s_{1}} = {\alpha _0}{\tilde s_{0}} + {\alpha _1}{\tilde s_{1}}$ holds, a pair of constellation ${{\mathcal S}_{0}}$ and ${{\mathcal S}_{1}}$ form an UDCP.
\label{def2}
\end{definition}
\vspace{0.1in}

Note that the UDCP can be considered as a special case of Definition \ref{def1} when ${\alpha _0}={\alpha _1}=1$. Definition \ref{def2} illustrates that if two constellations satisfy Definition \ref{def1}, after an amplitude transformation, the two constellations still meet Definition \ref{def1}. Therefore, we can extend Definition \ref{def2} to multiple constellations. Combining Definitions \ref{def1} and \ref{def2}, we give another definition as follows.

\vspace{0.1in}
\begin{definition}
Assume ${\alpha _n} \in A$, where $A = {\left[ {{a_0},{a_1}, \cdots ,{a_V}} \right]^T}, 0 < {a_0} < {a_1} <  \cdots  < {a_V}$, denotes a vector with real elements. If and only if ${s_n} = {\tilde s_n}$ such that $\sum _{n = 0}^N{\alpha _n}{s_n} = \sum _{n = 0}^N{\alpha _n}{\tilde s_n}$, we say that $\{ {{\mathcal S}_n}\} _{n = 0}^N$ form an UDCG.
\label{def3}
\end{definition}

\vspace{0.1in}
Let $c_n = {\alpha _n}{s_n}$ and ${\tilde c_n} = {\alpha _n}{\tilde s_n}$ in Definition \ref{def3}, and assume that $c_n \in {C}_n$ and $\tilde c_n \in { C}_n$. According to Definition \ref{def1}, we have Lemma \ref{lem1}.

\vspace{0.1in}
\begin{lemma}
If $\{ {{\mathcal S}_n}\} _{n = 0}^N$ forms an UDCG, then constellation group $\{ {{C}_n}\} _{n = 0}^N$ forms an UDCG, where ${ C_n} = A{{\mathcal S}_n}$.
\label{lem1}
\end{lemma}

\vspace{0.1in}
Here, Lemma \ref{lem1} explains that if a constellation group meets the UDCG constraint, the constellation group after the amplitude transformation still meets the UDCG constraint. If an UDCG can be found, after different amplitude transformations, we can obtain multiple constellation groups satisfying the UDCG constraint.

\subsection{Proposed Uniquely Decomposable Constellation Group}
According to Definition \ref{def3} and Lemma \ref{lem1}, we propose an UDCG in this subsection, which will be used for the SCMA codebook design in the next subsection. The proposed constellation group can be divided into two propositions as follows.
\vspace{0.1in}
\begin{proposition}
If $A = {\left[ {{a_0},{a_1}, \cdots ,{a_V}} \right]^T}$, $V = \frac{M}{{{2^r}}} - 1$ and $\theta  =  \pm \frac{\pi }{{{2^{r + \varepsilon}}}}$, where $0 < {a_0} < {a_1} <  \cdots  < {a_V}$, $\varepsilon > 0$, and $r$ is a positive integer satisfying $r \leqslant \log 2\left( M \right)$, then ${ C_0} = A{e^{j\frac{{2\pi }}{{{2^r}}}l}}$, ${ C_1} = {e^{j{\frac{\pi }{{{2^{r + \varepsilon}}}}}}}{ C_0} = A{e^{j\left( {\frac{{2\pi }}{{{2^r}}}l + \frac{\pi }{{{2^{r + \varepsilon}}}}} \right)}}$, and ${ C_2} = {e^{-j{\frac{\pi }{{{2^{r + \varepsilon}}}}}}}{ C_0} = A{e^{j\left( {\frac{{2\pi }}{{{2^r}}}l - \frac{\pi }{{{2^{r + \varepsilon}}}}} \right)}}$ form an UDCG, where $l \in \left\{ {0,1, \cdots ,{2^r} - 1} \right\}$.
\label{propos1}
\end{proposition}

\vspace{0.1in}
\emph {Proof:} See Appendix \ref{appA}.

\begin{figure}[t]
\centering
\subfigure[Phase rotation angles $\theta_1$ and $\theta_2\ (\theta_1 = -\theta_2)$ of base constellation with $M=8$ in Proposition \ref{propos1}.]{
\includegraphics[width=0.21\textwidth]{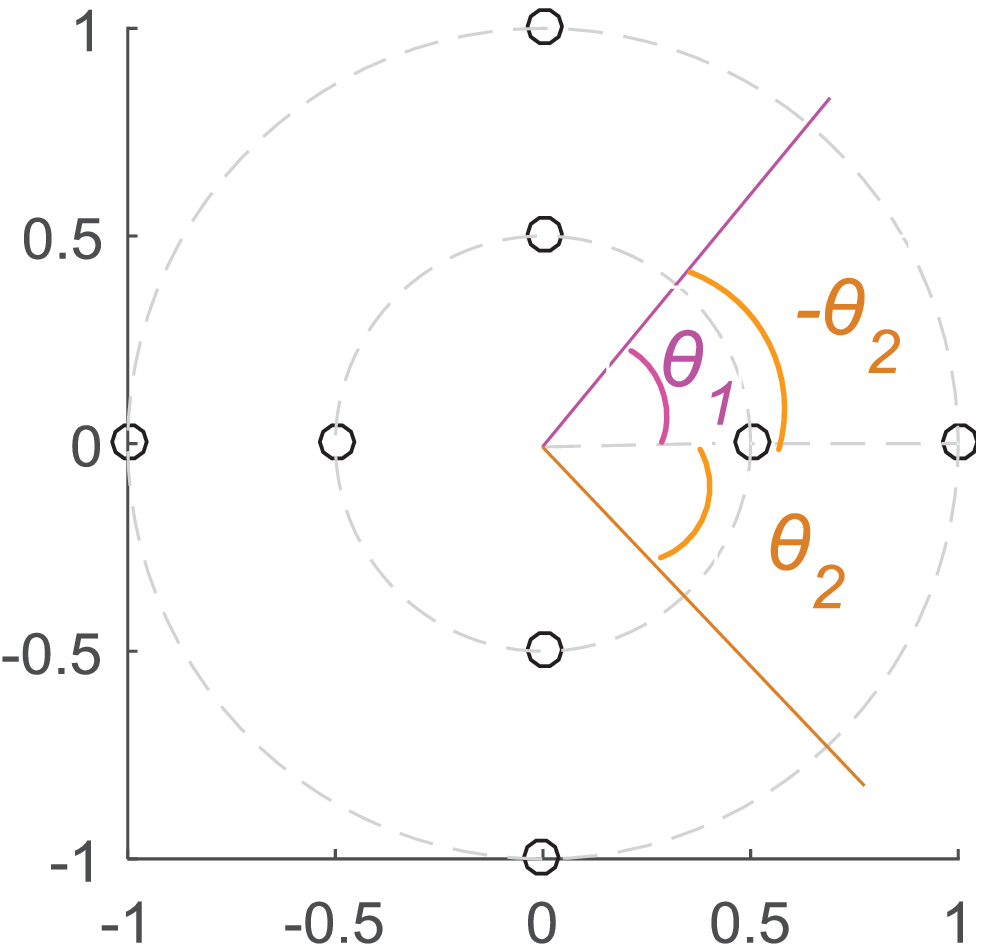}
\label{Fig:propos1} }
\hspace{2mm}
\subfigure[Phase rotation angles $\theta_1$ and $\theta_2\ (\theta_1 \neq -\theta_2)$ of base constellation with $M=8$ in Proposition \ref{propos2}.]{
\includegraphics[width=0.21\textwidth]{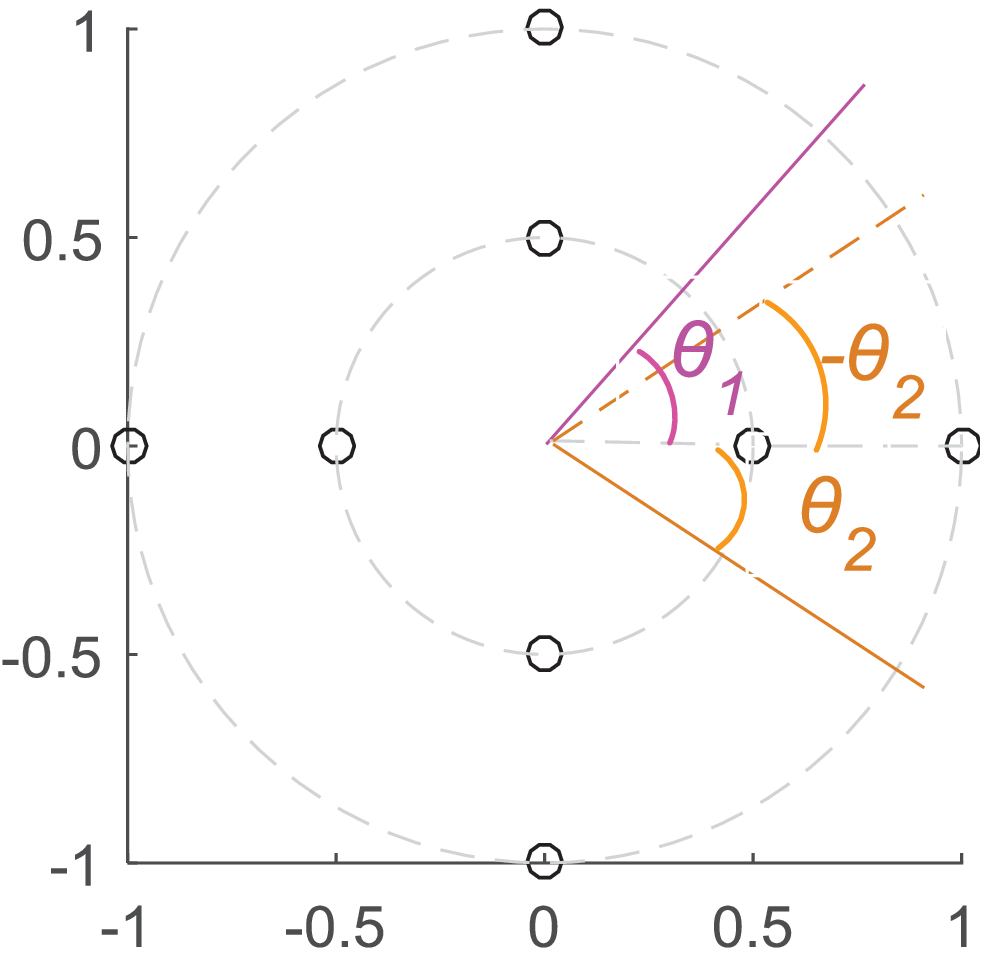}
\label{Fig:propos2}}
\label{myfigure}
\caption{Phase rotation constellations generated from the base constellations with $M=8$ in Propositions \ref{propos1} and \ref{propos2}.}
\end{figure}

\vspace{0.1in}
Giving a base constellation $ C_0$ and phase rotation constellations $\{ {{ C}_n}\} _{n = 1}^N$, Proposition \ref{propos1} illustrates that the base constellation and the phase rotation constellations form an UDCG, as shown in Fig. \ref{Fig:propos1}, where the phase rotation constellations can be obtained by the base constellation with phase rotations $\theta_1$ and $\theta_2\ (\theta_1 = -\theta_2=\frac{\pi }{{{2^{r + \varepsilon}}}})$. When $\theta_1 \neq -\theta_2$, as shown in Fig. \ref{Fig:propos2}, we then have Proposition \ref{propos2} as follows.

\vspace{0.1in}
\begin{proposition}
If $A = {\left[ {{a_0},{a_1}, \cdots ,{a_V}} \right]^T}$, $V = \frac{M}{{{2^r}}} - 1$, ${\theta _1} =  \pm \frac{\pi }{{{2^{r + {\varepsilon_1}}}}}$ and ${\theta _2} =  \pm \frac{\pi }{{{2^{r + {\varepsilon_2}}}}}$, where $0 < {a_0} < {a_1} <  \cdots  < {a_V}$, ${\varepsilon_1} \ne {\varepsilon_2} \geqslant 0$, and $r$ is a positive integer satisfying $r \leqslant \log 2\left( M \right)$, then ${ C_1} = A{e^{j\left( {\frac{{2\pi }}{{{2^r}}}l \pm \frac{\pi }{{{2^{r + {\varepsilon_1}}}}}} \right)}}$ and ${ C_2} = A{e^{j\left( {\frac{{2\pi }}{{{2^r}}}l \pm \frac{\pi }{{{2^{r + {\varepsilon_2}}}}}} \right)}}$ form an UDCG, where $l \in \left\{ {0,1, \cdots ,{2^r} - 1} \right\}$.
\label{propos2}
\end{proposition}

\vspace{0.1in}
\emph{Proof:} See Appendix B.

\vspace{0.1in}
Combining Propositions \ref{propos1} and \ref{propos2}, we can derive Lemma \ref{lem3} as follows.

\vspace{0.1in}
\begin{lemma}
There exists a constellation group containing constellations $\{ {{ C}_n}\} _{n = 0}^N$ to form an UDCG, and each constellation ${ C}_n$ contains $M$ constellation points. The UDCG can be expressed as
\begin{equation} \notag
{ C_n} =\left\{
\begin{aligned}
&{ C_0} = A{e^{j\frac{{2\pi }}{{{2^r}}}l}}, ~~ n = 0,\\
&{ C_n} = {e^{j{\theta _n}}}{ C_0}, ~~ n \geqslant 0,
\end{aligned}
\right.
\end{equation}
where $A = {\left[ {{a_0},{a_1}, \cdots ,a{}_V} \right]^T}$, $0 < {a_0} < {a_1} <  \cdots  < {a_V}$, $V = \frac{M}{{{2^r}}} - 1$, ${\theta _0}=0$, ${\theta _n} =  {{( - 1)}^{n}}\frac{\pi }{{{2^{r + {\varepsilon _n}}}}},  n > 0$, $0 < {\varepsilon_1} = {\varepsilon_2} <  \cdots  < {\varepsilon_{n-1}} = {\varepsilon_{n}}$, $l \in \left\{ {0,1, \cdots ,{2^r} - 1} \right\}$, and $r \leqslant \log 2\left( M \right)$ is a positive integer.
\label{lem3}
\end{lemma}

\vspace{0.1in}
From Lemma \ref{lem3}, there are $N+1$ constellations to form an UDCG. The constellation group consists of one base constellation $ C_0$ and $N$ phase rotation constellations $\{ {{ C}_n}\} _{n = 1}^N$.

\subsection{UDCG-based SCMA codebook design}
In this subsection, we propose a SCMA codebook design scheme using the UDCG as proposed in Subsection III.B. We allocate the constellations in UDCG to $d_f$ UNs. According to \cite{zhou2017scma}, we know that a codebook set with Latin feature can improve the BER performance. A codebook set with Latin feature satisfies $x_{{k_1},j}^m \ne x_{{k_2},j}^m$ when ${k_1} \ne {k_2}$, $x_{{k_1},m_j} \ne 0$ and $x_{{k_2},m_j} \ne 0$, which indicates that the non-zero elements in each codeword are different. Therefore, we can formulate a constellation allocation rule as
\begin{equation}\label{eq:c_inedx}
n _{kj} = (k + {u_j})\bmod ({d_f}),
\end{equation}
where $n _{kj}$ denotes the constellation index for the $j$-th UN on the $k$-th RN, and $1 \leqslant {u_j} \leqslant {d_f}$, which can be calculated as
\begin{equation}
{u_j} = \sum\limits_{i = 0}^j {{f_{k,i}}} ,\ j \in {\xi _k}.
\end{equation}
where $f_{k,j}$ denotes the element of ${\bm f^k} = \left[ {{f_{k,1}}, \cdots ,{f_{k,J}}} \right]$ for the $j$-th UN, and ${\bm f^k}$ represents the $k$-th row of $\bm F$, whose design methods were proposed in \cite{zhou2017scma} and \cite{peng2017joint}. Compared to \cite{peng2017joint}, the proposed $\bm F$ design method in \cite{zhou2017scma} is convenient, which is used to generate $\bm F$ in this paper. The $\bm F$ design scheme can be summarized as
\begin{equation}
D({\bm {f}_j}) = \sum\limits_{k = 1}^K {f_{k,j} {2^{K - k}}},
\label{eq:F_design}
\end{equation}
where $D({\bm {f}_j})$ denotes a binary-to-decimal mapping from $\bm {f}_j = \left[ {{f_{1,j}}, \cdots ,{f_{K,j}}} \right]$ and $\{ {D({\bm f_j})} \}_{j = 1}^J$ satisfies $D({\bm {f}_1})> \cdots > D({\bm {f}_J})$. With Eq. (\ref{eq:c_inedx}), a codebook generation matrix $\bm G$ can be obtained. For instance, a codebook generation matrix $\bm G_{4 \times 6}$ based on Fig. \ref{UNvsRN} can be written as
\begin{align}
\bm G_{4 \times 6}=\begin{bmatrix}
{ C_2} & { C_0} &{ C_1} & 0     &0       &0  \\
{ C_0} & 0      &0      &{ C_1} &{C_2}   &0 \\
0      & { C_1} &0      &{C_2}  &0       &{C_0}   \\
0      &0       &{ C_2} &0      &{ C_0}  & {C_1}
\end{bmatrix}.
\label{eq:GCB}
\end{align}
According to the generation matrix $\bm G$, the $j$-th UN's codebook set can be easily obtained by
\begin{equation}
{\mathcal {\bm X}_j} = {\bm f_j}  \circ  {\bm g_j},
\label{eq:CB}
\end{equation}
where $ \circ $ represents Hadamard product and ${\bm g_j}$ denotes the $j$-th column in $\bm G$.

\begin{figure}[t]
\centering
\subfigure[Constellations $C_0$, $C_1$ and $C_2$ with $r=3$ in $N=2$ $M=8$ UDCG.]{
\includegraphics[width=0.5\textwidth]{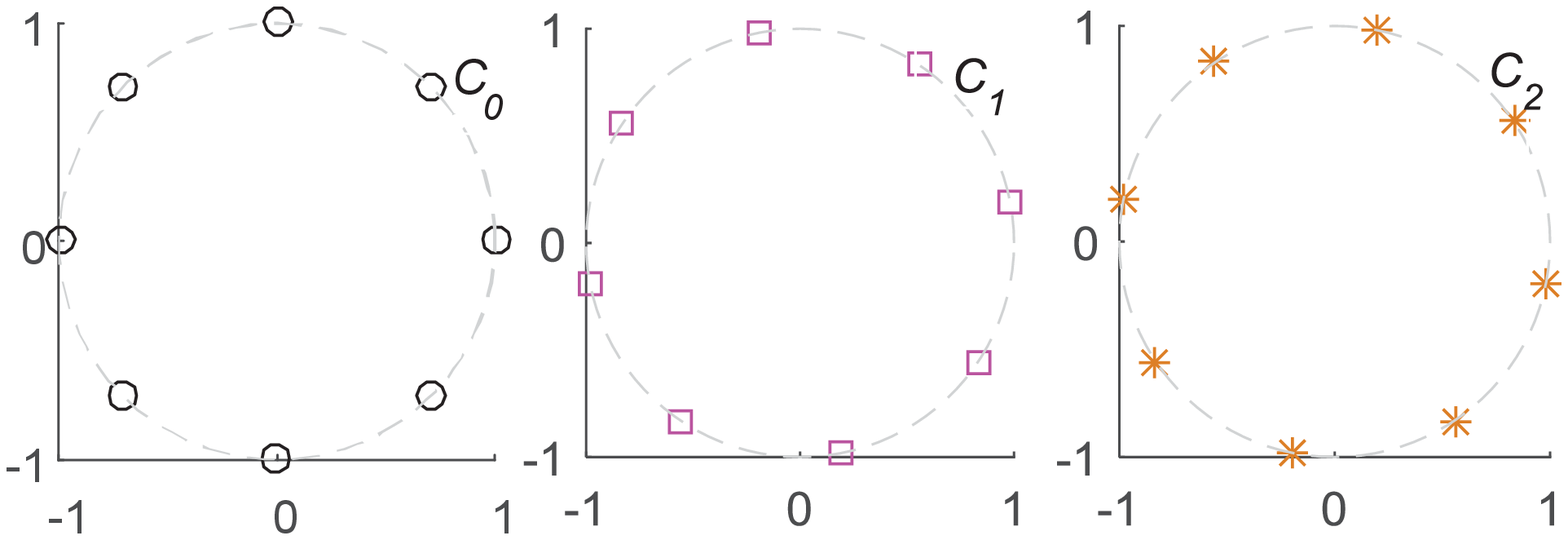}
\label{Fig:case1} }
\vspace{0.125in}
\subfigure[Constellations $C_0$, $C_1$ and $C_2$ with $r=2$ in $N=2$ $M=8$ UDCG.]{
\includegraphics[width=0.5\textwidth]{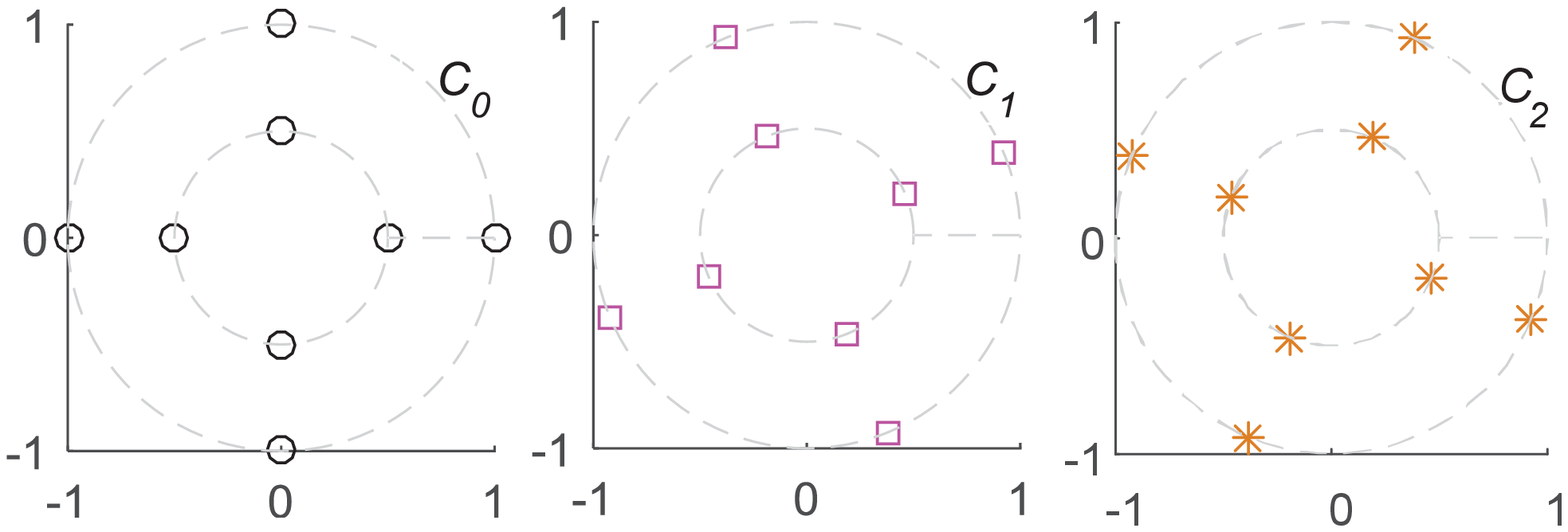}
\label{Fig:case2}}
\label{}
\caption{Two cases of constellations with $M=8$ in $N=d_f-1=2$ UDCG.}
\end{figure}

It is shown in (\ref{eq:CB}) that the proposed codebook design scheme has a relatively low computation complexity compared to the method in \cite{zhang2019efficient}. It needs only to allocate the constellations contained in the UDCG to different UNs on the same RN according to (\ref{eq:c_inedx}). However, there is an implementation issue when $M$ is large. This issue is illustrated with an example as shown in Fig. \ref{Fig:case1} and Fig. \ref{Fig:case2}, which represent two cases of constellations with $M=8$ in $N=d_f-1=2$ UDCG. When $r=3$, all constellation points have the same magnitude $A = {a_0}$, as shown in Fig. \ref{Fig:case1}. When $r=2$, there will be two amplitudes $A = \left[ {a_0},{a_1} \right]^T$, as shown in Fig. \ref{Fig:case2}. However, in the second case, we do not know on what scale the interval between $a_0$ and $a_1$ should be to satisfy a desired BER, which will be discussed in the following section.

\vspace{0.125in}
\section{Implementation issues}
Next, we will discuss about the implementation issues as mentioned in the previous section. For discussion convenience, we assume that $a_0=a$ denotes the base length and $a_v=a+vt$ in $A$, where $v \in[ {0,V}]$ and $t$ is a step size. The implementation issues can be described as how large $t$ should be to ensure that the UDCG-based codebooks can offer optimal BER performance. According to previous studies in \cite{zhang2019efficient,8688492,29611}, the coding gain ${{d_{\min }^2} \mathord{\left/{\vphantom {{d_{\min }^2} {\bar E}}} \right. \kern-\nulldelimiterspace} {\bar E}}$ is a key performance indicator (KPI) for constellation design, where ${\bar E}$ is the average energy of the constellation. In addition, the data of each UN is mapped into multiple RNs through a codebook set for transmission. We can find the optimal step size $t$ from two perspectives, i.e., in terms of single RN and multiple RNs. From single RN perspective, an optimal step size $t_c$ is found to maximize the coding gain of superimposed constellation. While the codebook sets' coding gain is considered to find the optimal $t_x$ from a multiple RNs perspective.

\subsection{Single Resource Node}
Let ${ C_s} = [ {{c_1}, \cdots ,{c_p}, \cdots ,{c_{{M^{{d_f}}}}}} ]$ denote the superimposed constellation $\uplus _{n = 0}^{d_f-1}{{ C}_n}$. From a single RN perspective, the problem to find optimal $t_c$ can be solved by maximizing the coding gain of the superimposed constellation $ C_s$ with a single RN, which is expressed as
\begin{equation}
{t_c }  = \mathop {\arg \max }\limits_{a, M, r} \frac{{\left({d_{\min }^c}\right)^2}}{\bar E_c},
\label{eq:opcase1}
\end{equation}
where the average energy ${\bar E_c}$ and the MED ${d_{\min }^c}$ of $ C_s$ are calculated respectively as
\begin{equation}
{{\bar E}_c} = \frac{1}{{{M^{{d_f}}}}}{E_c} = \frac{1}{{{M^{{d_f}}}}}\sum\limits_{p = 1}^{{M^{{d_f}}}} {{{\left\| {{c_p}} \right\|}^2}},
\label{eq:barEc}
\end{equation}
\begin{equation}
~~~~~~~~~~~{d_{\min }^c} = \mathop {\min }\limits_{{p_1} \ne {p_2}} \left( {\left\| {{c_{{p_1}}} - {c_{{p_2}}}} \right\|} \right),~~{c_{{p_1}}},{c_{{p_2}}} \in { C_s},
\label{eq:dminc}
\end{equation}
where ${{ E}_c}$ denotes the sum energy. According to Lemma \ref{lem3}, $E_c$ can be calculated as
\begin{equation}
\begin{aligned}
{E_c} = & {d_f}{M^{({d_f} - 1)}}{2^r}\bigg[ {\sum\limits_{v = 0}^V {{{\left( {a + vt} \right)}^2}} } \bigg] + \\
& 2{M^{({d_f} - 2)}} \bigg[ {\sum\limits_{q = 1}^{{2^r}} {2q\cos \left( {\frac{{2\pi q}}{{{2^r}}}} \right) - {2^r}} } \bigg]  \\
&  \times {\bigg[ {\sum\limits_{v = 0}^V {\left( {a + vt} \right)} } \bigg]^2}\bigg[ {\sum\limits_{n = 0}^{{d_f} - 1} {\sum\limits_{i > n}^{{d_f} - 1} {\cos \left( {{\theta _n} - {\theta _i}} \right)} } } \bigg].
\end{aligned}
\label{eq:Ec}
\end{equation}
The derivation of $E_c$ is given in Appendix \ref{appC}.

Moreover, two square distances can be derived from $d_c^2 ={\left\| {{c_{{p_1}}} - {c_{{p_2}}}} \right\|^2}$. Assume that $\Delta _\theta ^n$ and $\Delta _a^n$ denote the angle difference and amplitude difference between two constellation points in the $n$-th constellation, respectively. When $2^r \leqslant M$, $\Delta _a^n=0, \forall n$ and ${\alpha _{{n}}}=a$, we consider only $\Delta _\theta ^{n} \ne 0$ and $\Delta _\theta ^{i} \ne 0, {i} \ne {n}$ for the $n$-th and $i$-th constellations, respectively. In this case, the minimum square distance can be given as
\begin{equation}
\begin{aligned}
{\left( {d_{1,\min }^c} \right)^2} = 8{a^2}{\left( {\sin \frac{\pi }
{{{2^r}}}} \right)^2}\left( {1 - \cos \theta _{1,\min }^c} \right),
\end{aligned}
\label{eq:dcMeqa}
\end{equation}
where ${\theta _{1,\min }^c} = \min[ {| {\frac{{2\pi }}{{{2^r}}}( {{l_n} - {l_i} - 1} ) + {\theta _n} - {\theta _i}}|} ]$ with $l_n, l_i \in  \left\{ 0,1, \cdots ,{2^r} - 1 \right\}$.

In addition, let us consider $\Delta _\theta ^n=0, \forall n$, $\Delta _a^{n} \ne 0$ for two constellations when $2^r<M$. In this case, the minimum square distance is expressed as
\begin{equation}
\left( d_{2,\min}^c \right)^2 = 2{t^2}\left( {1 - \cos {\theta _{2,\min }^c}} \right),
\label{eq:dcMnea}
\end{equation}
where ${\theta _{{2,\min}}^c}= \min[ {| {\frac{{2\pi }}{{{2^r}}}( {{l_n} - {l_i}} ) + {\theta _n} - {\theta _i}} |} ]$.

The derivations of (\ref{eq:dcMeqa}) and (\ref{eq:dcMnea}) are given in Appendix \ref{appD}. The distance ${d_{1,\min }^c}$ is a constant giving the base length $a$, while ${d_{2,\min }^c}$ increases with $t$ increasing. Therefore, it can be inferred that (\ref{eq:opcase1}) increases first and then decreases with $t$ increasing. As a consequence, we can adopt an exhaustive method to solve (\ref{eq:opcase1}) by pre-setting a search range $( {{t_{ini}},{t_{end}}} )$, $t_{ini}<t_{end}$, where $t \in ( {t_{ini},t_{end}})$. The range $( {t_{ini},t_{end}} )$ can be adjusted according to the value of $a$.

\subsection{Multiple Resource Nodes}
Let ${d_{\min }^x}$ and ${\bar E_x}$ denote the MED and the average energy of the codebook $\mathcal X_j$. From a multiple RNs perspective, the problem to find optimal $t_x$ can be solved by maximizing the coding gain of $\mathcal X_j$, which can be written as
\begin{equation}
{t_x }  = \mathop {\arg \max }\limits_{a, M, r} \frac{{\left({d_{\min }^x}\right)^2}}{\bar E_x},
\label{eq:opcase2}
\end{equation}
where
\begin{equation} \label{eq:barEx}
{{\bar E}_x} = \frac{1}{M}{E_x} = \frac{1}{M}\sum\limits_{m = 1}^M {{{\left\| {{\bm x_{m,j}}} \right\|}^2}},
\end{equation}
\begin{equation}
{d_{\min }^x} = \mathop {\min }\limits_{{m_1} \ne {m_2}} \left( {\left\| {{\bm x_{{m_1},j}} - {\bm x_{{m_2},j}}} \right\|} \right),  {\bm x_{{{m_1},j}}},{\bm x_{{{m_2},j}}} \in {\mathcal X_j},
\end{equation}
where $E_x$ denotes the sum energy of $\mathcal X_j$. According to (\ref{eq:GCB}), the expression of $E_x$ can be calculated as
\begin{equation} \label{eq:Ex}
{E_x} = {2^{r + 1}}\left[ {\sum\limits_{v = 0}^V {{{\left( {a + vt} \right)}^2}} } \right].
\end{equation}
Moreover, the MED among the codewords of each codebook is determined by step size $t$ and $2a\sin( {\frac{\pi }{{{2^r}}}})$, where $2a\sin( {\frac{\pi }{{{2^r}}}})$ denotes the distance between two adjacent points on the innermost circle. As a result, $d_{min}^x$ can be rewritten as
\begin{equation} \label{eq:dminx}
d_{min}^x=\left\{
\begin{array}{ll}
2\sqrt 2 a\sin \left( {\frac{\pi }{{{2^r}}}} \right),&t \geqslant 2a\sin \left( {\frac{{2\pi }}{{{2^{r + 1}}}}} \right), \\
\sqrt 2 t,&0 < t < 2a\sin \left( {\frac{{2\pi }}{{{2^{r + 1}}}}} \right), \\
2\sqrt 2 a\sin \left( {\frac{\pi }{{{2^r}}}} \right),&t = 0.
\end{array}
\right.
\end{equation}

According to (\ref{eq:Ex}) and (\ref{eq:dminx}), the objective function of (\ref{eq:opcase2}) increases first and then decreases as $t$ increases, which is similar to (\ref{eq:opcase1}), such that an exhaustive method can be adopted to solve it.

In addition, minimum product distance (MPD) $\mathop {\min }\limits_{{i_1},{i_2} \in {\mathcal X}} \prod\limits_{k = 1}^K {\left| {x_k^{({i_1})} - x_k^{({i_2})}} \right|} $ of $\mathcal X$ is a KPI for SCMA codebook design over RF channel \cite{8688492}. To further enlarge the MPD and improve the BER performance, a simple interleaving operation \cite{cai2016multi,yan2016top} is used to rearrange the order of constellation points of the constellation $C_n$ that is allocated to non-zero even rows of $\mathcal X_j$. Let $C_n^e$ denote the constellation of $C_n$ after interleaving, it can be expressed as
\begin{equation}
C_n^e = {e^{j{\theta _n}}}{C_0^I}, ~\text{when}~{\mu ^k}\bmod 2 =0,
\label{eq:inl}
\end{equation}
with
\begin{equation}
C_0^I = {A^d}{C_0}\left( {I_d} \right),~{\mu ^k} = \sum\limits_{i = 0}^k {{f_{i,j}}}, k \in {\varsigma _j},
\label{}
\end{equation}
where ${A^d} = {\left[ {{a_V},{a_{V-1}}, \cdots ,{a_0}} \right]^T}$ denotes the inversion of $A$, ${\varsigma _j}$ represents the non-zero positions in $j$-th column of $F$, and $I_d$ denotes the constellation points indices of base constellation $C_0$ after interleaving. Let $M_r=2^r$ and $M_h=2^{r-1}$, we have ${I_d} = \left\{ {I_d^o,I_d^e} \right\}$ where odd indices $I_d^o = \left\{ {{M_h} + 1, \cdots ,{M_r} - 1,1, \cdots ,{M_h} - 1} \right\}$ and even indices $I_d^e = \left\{ {{M_h} + 2, \cdots ,{M_r},2, \cdots ,{M_h}} \right\}$. In this case, the UDCG-based codebooks can be generated using Algorithm \ref{alg1}.
\begin{algorithm}
\small
\caption{The UDCG-based codebook design procedures.}
\label{alg1}
\begin{algorithmic}
\STATE Initialize $a, M, r, \theta_n, t_{ini}, t_{end}$;
\STATE $G_c^{old}=G_x^{old}=0$; $t_c=t_x=0$;
\FOR{$t \in [t_{ini},t_{end}]$}

\IF {\text{Case~1}}
\STATE Calculate $d_{\min}^c$ by (\ref{eq:dminc});
\STATE Calculate $\bar E_c$ by (\ref{eq:barEc}) and (\ref{eq:Ec});
\STATE Calculate $G_c^{new}={\left( {d_{\min }^c} \right)^2}/{{\bar E}_c}$;
\IF{$G_c^{new}>G_c^{old}$}
\STATE $G_c^{old}=G_c^{new}$; $t_c=t$;
\ENDIF

\ELSE
\STATE Calculate $d_{\min}^x$ by (\ref{eq:dminx});
\STATE Calculate $\bar E_x$ by (\ref{eq:barEx}) and (\ref{eq:Ex});
\STATE Calculate $G_x^{new}={\left( {d_{\min }^x} \right)^2}/{{\bar E}_x}$;
\IF{$G_x^{new}>G_x^{old}$}
\STATE $G_x^{old}=G_x^{new}$; $t_x=t$;
\ENDIF
\ENDIF
\ENDFOR

\STATE Generate an UDCG with $N=d_f-1$ by $t_c$ or $t_x$ and Lemma \ref{lem3};
\STATE Obtain codebook $\mathcal X_j$ by (\ref{eq:c_inedx}), (\ref{eq:F_design}) and (\ref{eq:CB});
\STATE Enhance the minimum PD of $\mathcal X_j$ by (\ref{eq:inl});
\RETURN $\mathcal X_j$;
\end{algorithmic}
\end{algorithm}

\vspace{0.125in}
\section{Simulation results and discussions}
In this section, compared with TMQAM-based OCB \cite{taherzadeh2014scma}, CRCB \cite{zhou2017scma}, LCRCB \cite{gao2017low}, multi-dimensional LCB \cite{zhang2019efficient} and GAMCB for downlink \cite{mheich2018design}, the BER performance of the proposed UDCG-based codebook design scheme is evaluated over AWGN and RF channels, respectively. In these simulations, all codebooks' power of different design schemes has been normalized as one. If no otherwise specified, we default $\varepsilon=1$ for $d_f=3$ SCMA systems. We obtain the optimal step size $t_c$ and $t_x$ by setting $t \in ( {0,2} )$ with $a=0.5$, and list the MED $d_{k,\min}$ of superimposed constellations at the $k$-th RN in Tab.~\ref{tab1} and Tab.~\ref{tab2}. Note that four most significant digits are retained in these tables for notation simplicity.

\subsection{BER Performance of Uncoded SCMA systems}
\begin{figure*}[t]
\centering
\subfigure [$M=4$, AWGN channel.] {\label{fig11}
\includegraphics[width=0.24\textwidth]{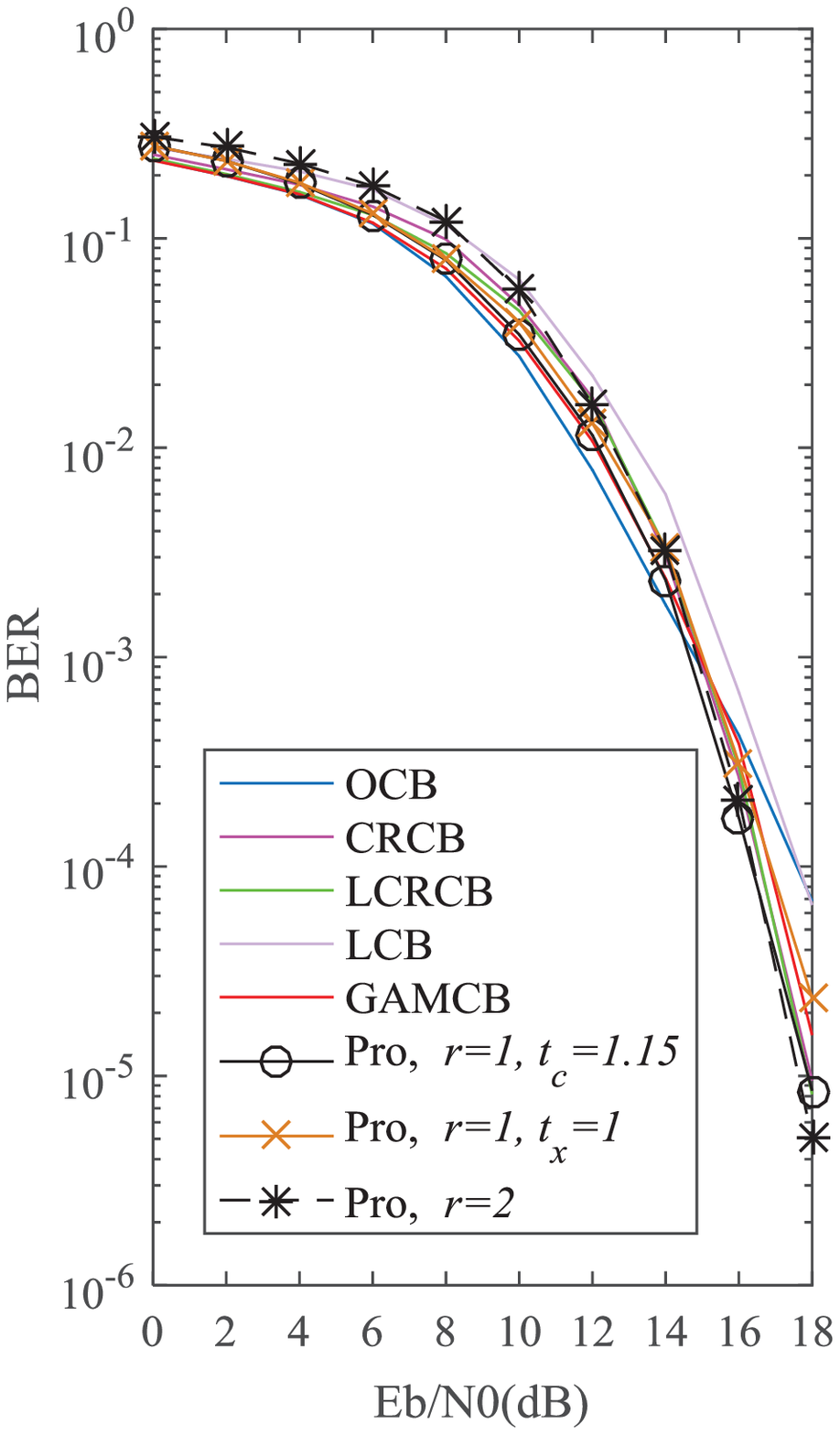}
}\hspace{8mm}
\subfigure[$M=4$, RF channel.] {\label{fig12}
\includegraphics[width=0.24\textwidth]{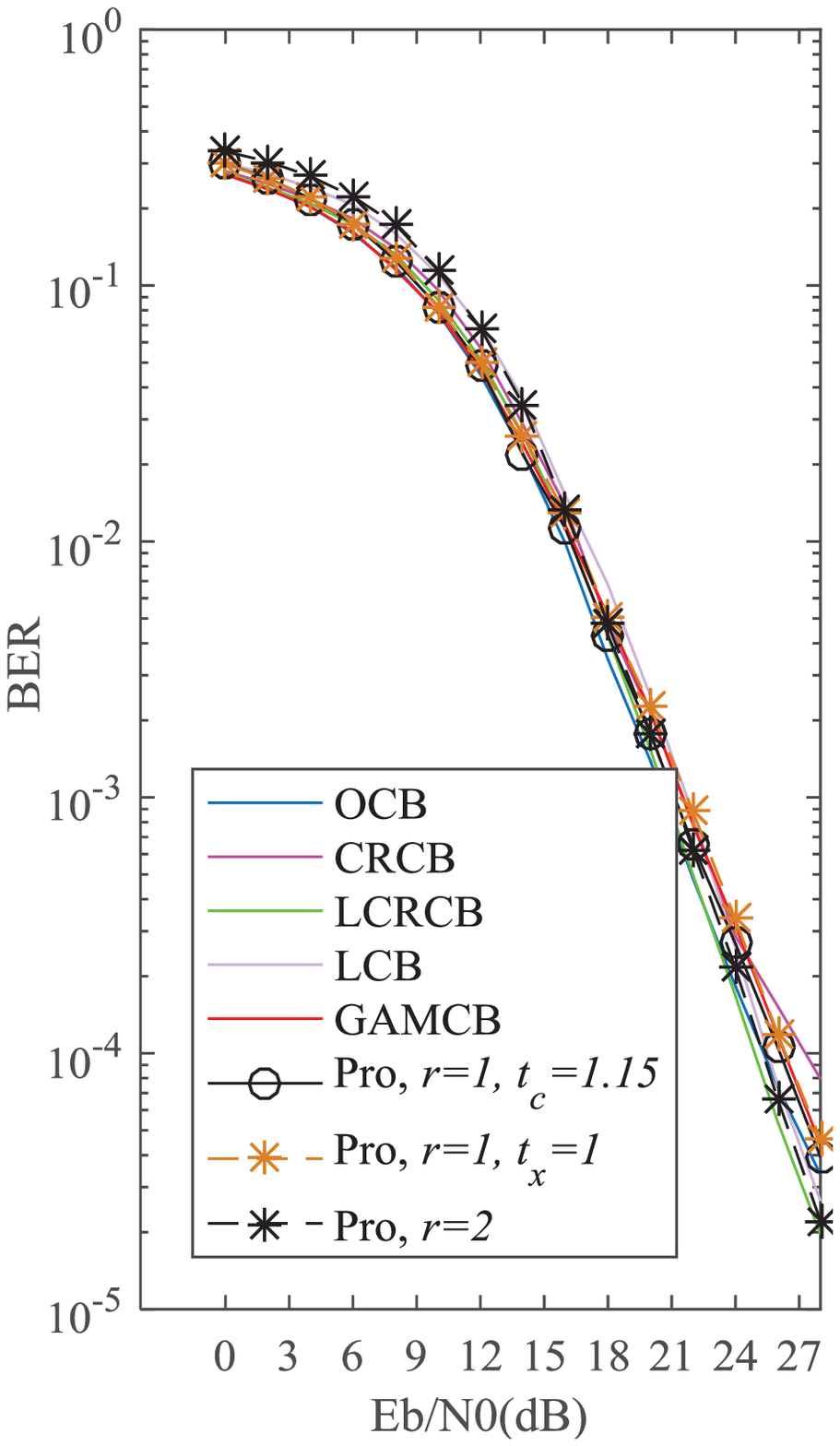}
}\hspace{8mm}
\subfigure[$M=8$, AWGN channel.]{ \label{fig21}
\includegraphics[width=0.24\textwidth]{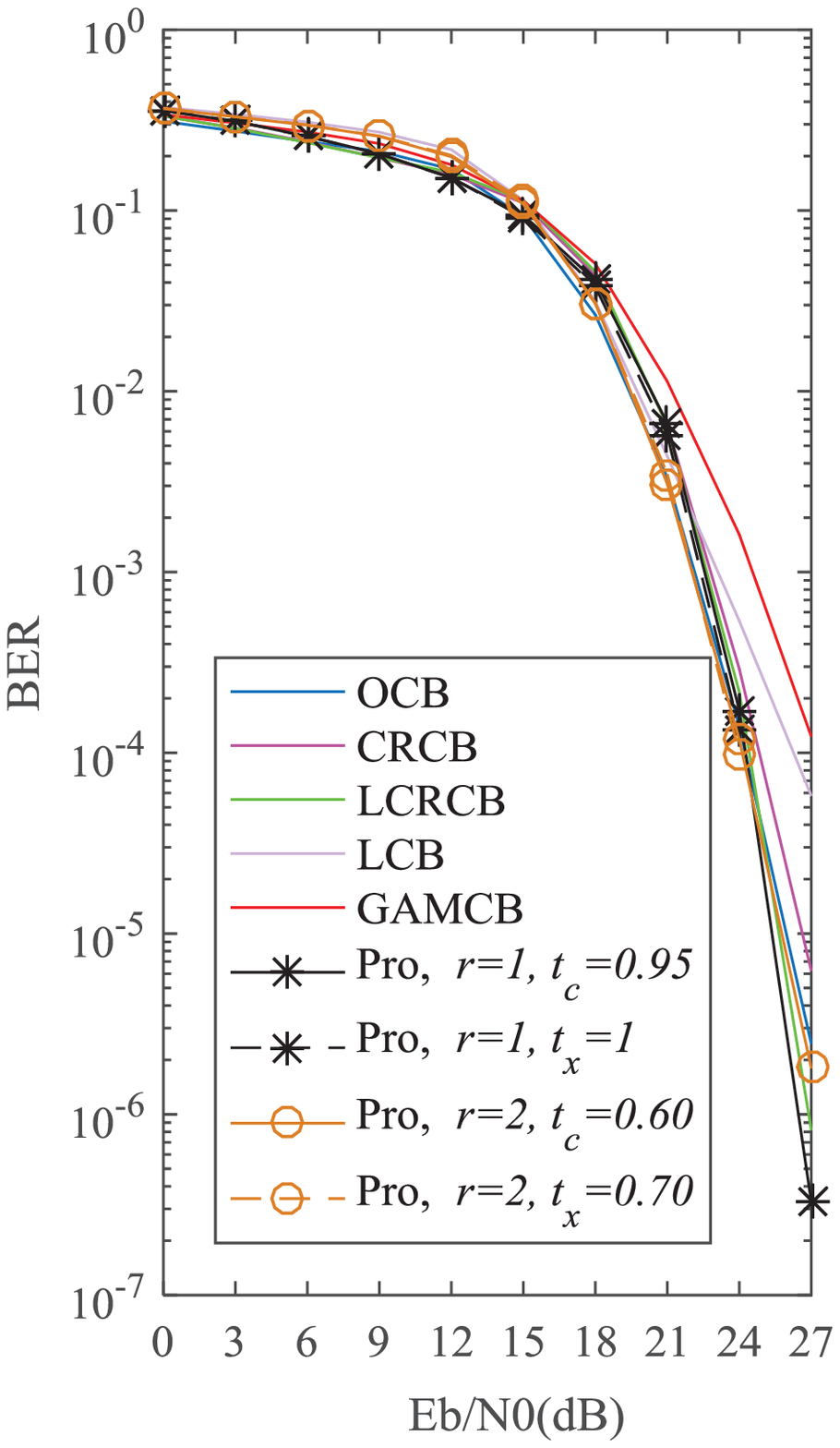}}
\hspace{8mm}  \\
\subfigure[ $M=8$, RF channel.]{\label{fig22}
\includegraphics[width=0.24\textwidth]{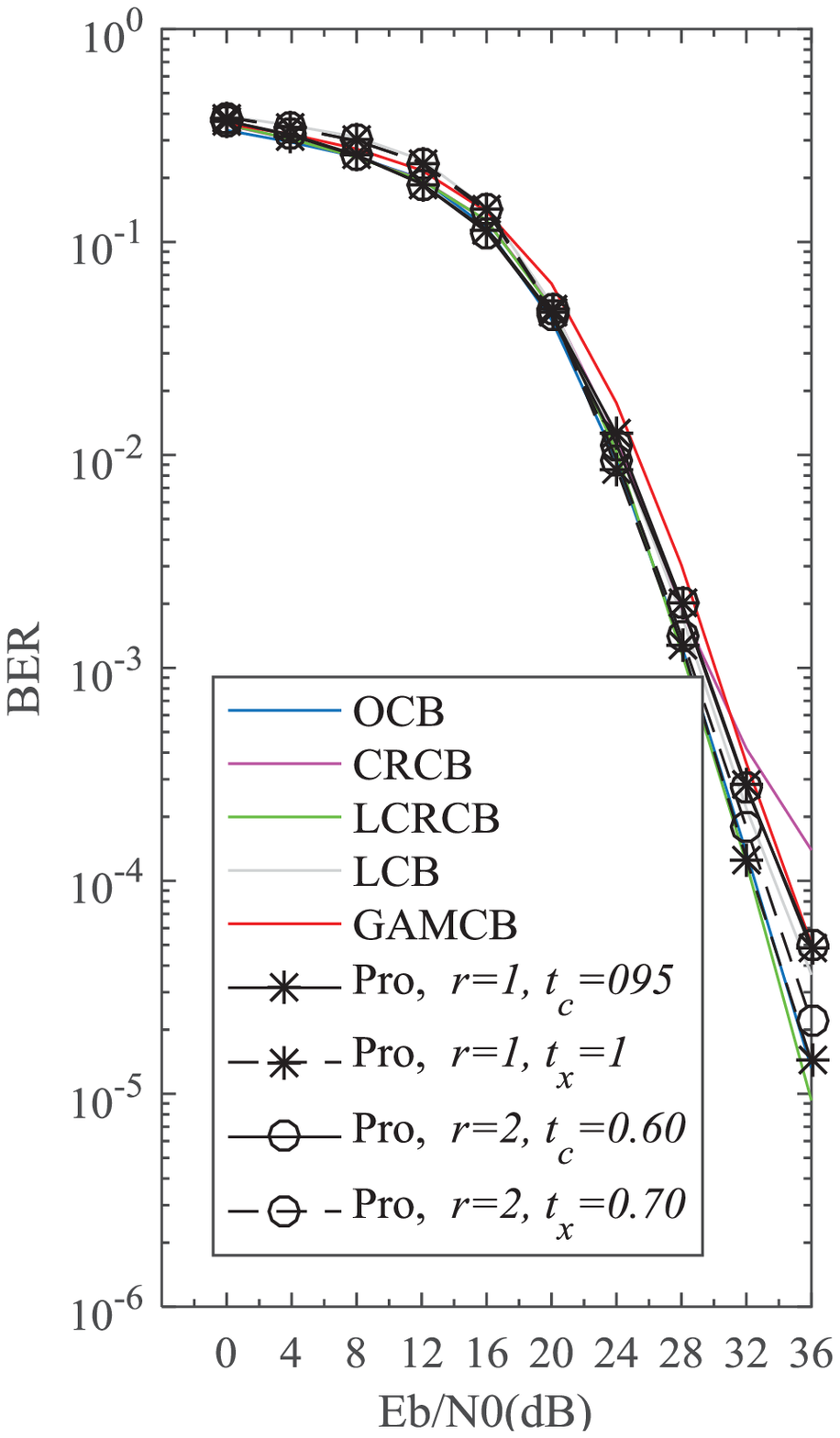}}
\hspace{8mm}
\subfigure[$M=16$, AWGN channel.]{ \label{fig31}
\includegraphics[width=0.24\textwidth]{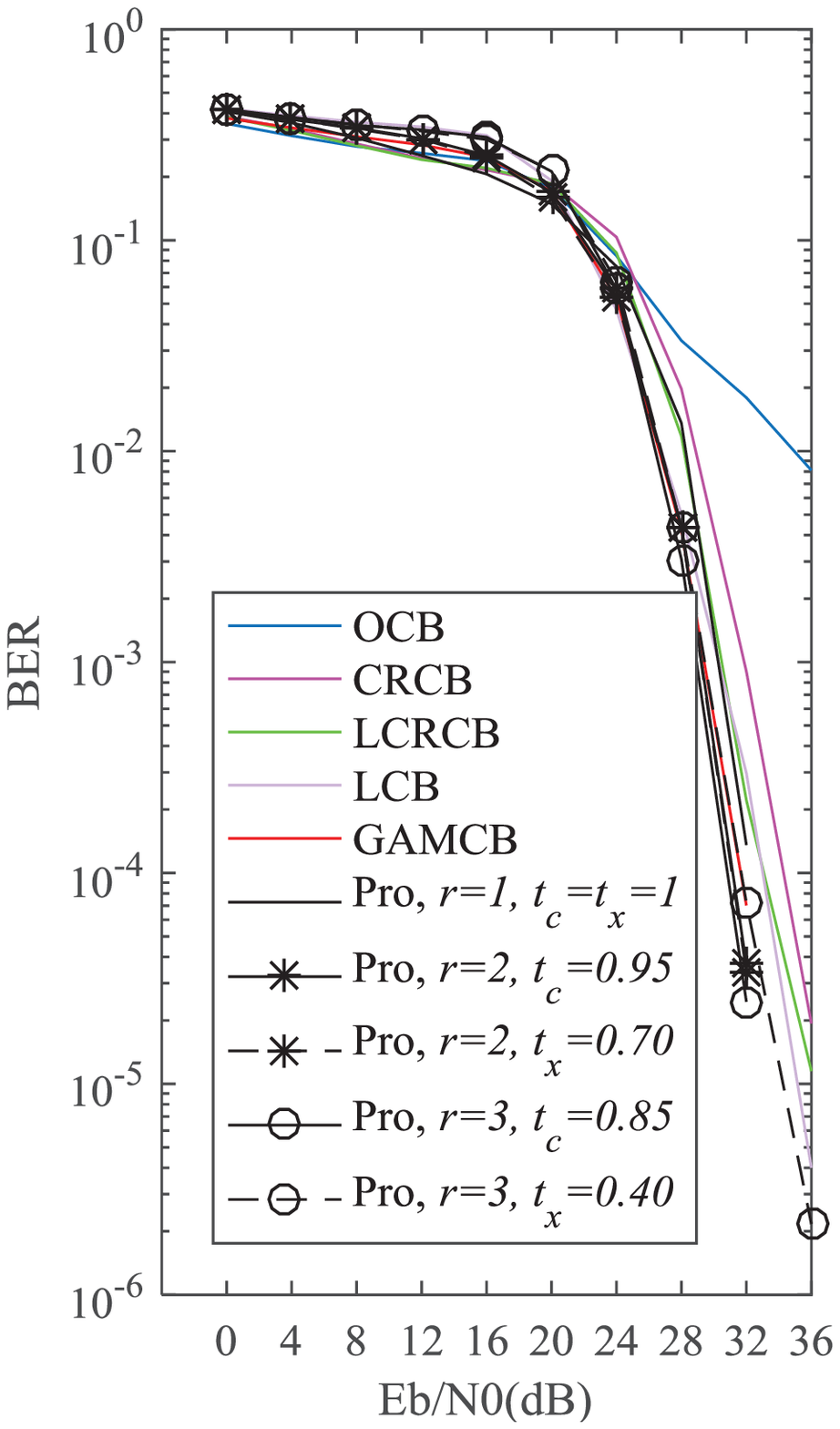}}
\hspace{8mm}
\subfigure[$M=16$, RF channel.]{\label{fig32}
\includegraphics[width=0.24\textwidth]{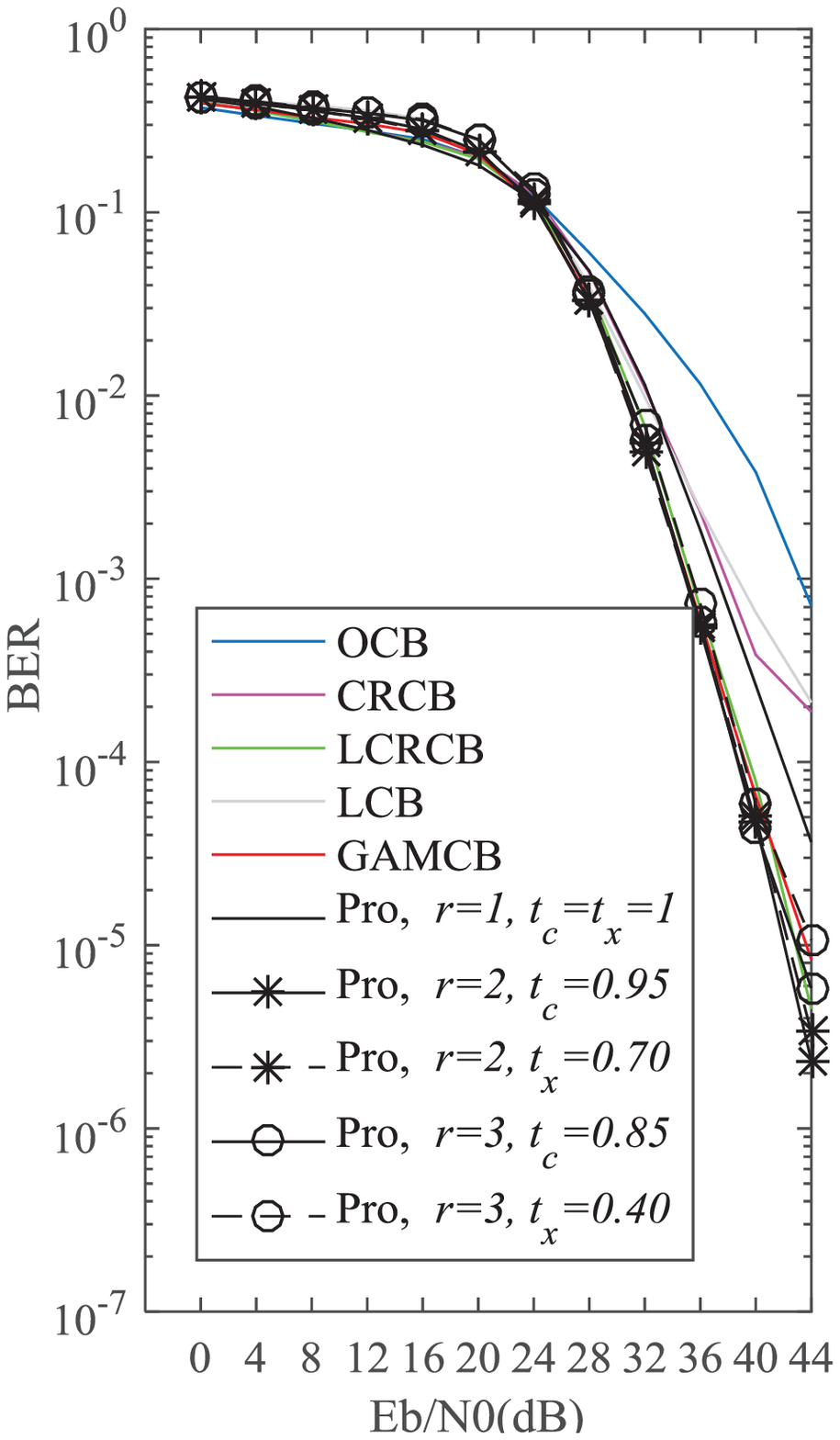}}
\caption{BER performance of uncoded SCMA systems with $M=4$, $M=8$ and $M=16$ over AWGN and RF channels.}
\label{Fig1}
\end{figure*}

In this subsection, we evaluate the BER performance of uncoded SCMA systems with $M=4$, $M=8$ and $M=16$ in Fig. \ref{Fig1}. We set $K=4$, $J=6$ and $d_f=3$ for these simulations. At the receiver side, MPA with $10$ iteration times is used as a multi-user detector.

Fig. \ref{fig11} and Fig. \ref{fig12} show the BER performance of different schemes when $M=4$ over AWGN and RF channels, respectively. From Fig. \ref{fig11}, we see that the BER performance of OCB is getting worse off in a high SNR region. The proposed scheme with $r=2$ has the worst BER performance with low Eb/N0, but the best BER performance with high Eb/N0. When BER = $10^{-4}$, the proposed codebooks with $r=1, t_c=1.15$ has extra gain 2 dB higher than OCB and LCB, and extra gain 0.5 dB higher than LCRCB. However, we see from Fig. \ref{fig12} that the BER performance of CRCB has 1 dB loss if compared to that of LCRCB when BER = $10^{- 4}$. The BER curves for LCB and OCB are close to the best possible curve, and the proposed design scheme still has good BER performance with high Eb/N0 over fading channel. That is because the proposed constellation with $d_{\min} \ne 0$ can further improve the MPA detection performance, especially with high Eb/N0.

Fig. \ref{fig21} and Fig. \ref{fig22} show the BER performance of different schemes with $M=8$ over AWGN and RF channels. We can see from these figures that the BER curve of OCB is close to that of LCRCB, and exceeds that of CRCB. And LCB has the worst BER performance over AWGN channel, but outperforms that of CRCB and the proposed scheme with $r=1, t_c=0.95$ and $r=2, t_c=0.60$ over RF channel. However, the BER performance of the proposed scheme with $r=1, t_x=1$ and $r=2, t_x=0.70$ can reach zero when Eb/N0 = 24 dB in Fig. \ref{fig21}, and is close to the best curve of LCRCB in Fig. \ref{fig22}. In addition, the proposed scheme with $r=1$ has better BER performance than that of the proposed scheme with $r=2$ in low Eb/N0 regions. When $r$ is the same, the BER performance of the proposed scheme with $t_x$ is better than that of the proposed scheme with $t_c$ in low Eb/N0 regions, but this advantage will disappear in high Eb/N0 regions.

Fig. \ref{fig31} and Fig. \ref{fig32} show the BER performance of different schemes with $M=16$ over AWGN and RF channels. We can see that the BER performance of OCB is the worst in both Figs. \ref{fig31} and \ref{fig32}. Although LCB has extra 2 dB gain than CRCB over AWGN channel when BER = $10^{-4}$, both of them have a similar BER performance over RF channel. In Fig. \ref{fig31}, the BER performance of the proposed scheme with $r=3, t_x=0.40$ is close to that of GAMCB, and going down to $10^{-6}$ at Eb/N0=36 dB. In particular, the proposed scheme with $r=2$ and $r=3, t_c=0.85$ can reach to zero when Eb/N0=36 dB. When BER = $10^{-5}$, the proposed scheme with $r=2$ has extra 2 dB gain than GAMCB, and extra 4 dB gain than LCRCB over AWGN channel. However, this advantage of the proposed scheme will be weakened over RF channel. In Fig. \ref{fig32}, GAMCB and LCRCB have the similar BER performance. Compared to GAMCB and LCRCB, the BER performance of the proposed scheme with $r=2$ is slightly better than that of them when BER = $10^{-5}$.

In addition, Tab. \ref{tab1} lists the MED $d_{k,\min}, k=1,\cdots, K$ of superimposed constellation, i.e, $d_{1,\min}/ \cdots / d_{K,\min}$. In Tab. \ref{tab1}, the proposed schemes with $r=2$ in Fig. \ref{fig11}, $r=2, t_x=0.70$ in Fig. \ref{fig21}, and $r=3, t_c=0.85$ in Fig. \ref{fig31} are used as statistical objects. According to Tab. \ref{tab1}, we see that when $M=4$, the MED $d_{k,\min}$ of OCB scheme is the largest one among all design schemes, and LCB and GAMCB's $d_{k,\min}$ is not always zero. With $M$ increasing, their $d_{k,\min}$ values are decreased. In particular, except for the proposed UDCG-based and GAMCB design schemes, the other design schemes' $d_{k,\min}$ are all close to zero. However, the MED $d_{k,\min}$ of the proposed scheme is not zero on each RN. Combining Fig. \ref{Fig1}, it can be derived that ${d_{\min }} = \mathop {\min }\limits_{k \in \left\{ {1, \cdots ,K} \right\}} \left( {{d_{k,\min }}} \right)$ is an important KPI for SCMA codebook design. When designing codebooks, $d_{\min}$ needs to be taken into consideration in order to achieve a desirable BER target.

\begin{table*}[t]
\centering
\caption{$d_{k,\min}$ of different design schemes with $M=4$, $M=8$ and $M=16$.}\label{tab1}
\begin{tabular}{lccc}
\hline
Category & {$M=4 \left( {d_{k,\min }} \right)$} & {$M=8 \left( {d_{k,\min }} \right)$} & {$M=16 \left( {d_{k,\min }} \right)$} \\
\hline
{OCB} & 0.0733/0.0732/0.0732/0.0733 &0.0010/0.0000/0.0000/0.0010 & 0.0000/0.0000/0.0000/0.0000 \\
{CRCB} &0.0000/0.0000/0.0000/0.0000 &0.0000/0.0000/0.0000/0.0000 & 0.0000/0.0000/0.0000/0.0000 \\
{LCRCB}&0.0000/0.0000/0.0000/0.0000 &0.0000/0.0000/0.0000/0.0000 & 0.0000/0.0000/0.0000/0.0000 \\
{LCB} &0.0000/0.1305/0.1305/0.0000 &0.0000/0.0000/0.0029/0.0032  &0.0000/0.0000/0.0000/0.0000 \\
{GAMCB} &0.0000/0.0438/0.0438/0.0000 &0.0003/0.0057/0.0043/0.0025&0.0003/0.0001/0.0002/0.0003  \\
{Proposed}&0.0412/0.0412/0.0412/0.0412 & 0.0008/0.0008/0.0008/0.0008 & 0.0012/0.0012/0.0012/0.0012 \\
\hline
\end{tabular}
\end{table*}

\begin{figure*}[t]
\centering
	\subfigure[BER performance of the proposed scheme with different $\varepsilon$ over AWGN channel, when $M=4$ and $M=8$.]{ \label{Fig4}  %
    \includegraphics[width=0.48\textwidth]{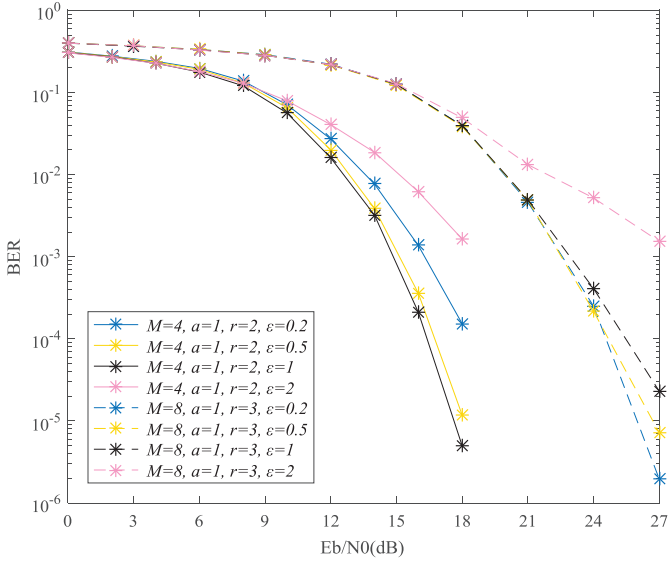}}
    \hspace{2mm}
	\subfigure[BER performance of optimal $t_c$ over AWGN channel.]{\label{Fig5}
    \includegraphics[width=0.48\textwidth]{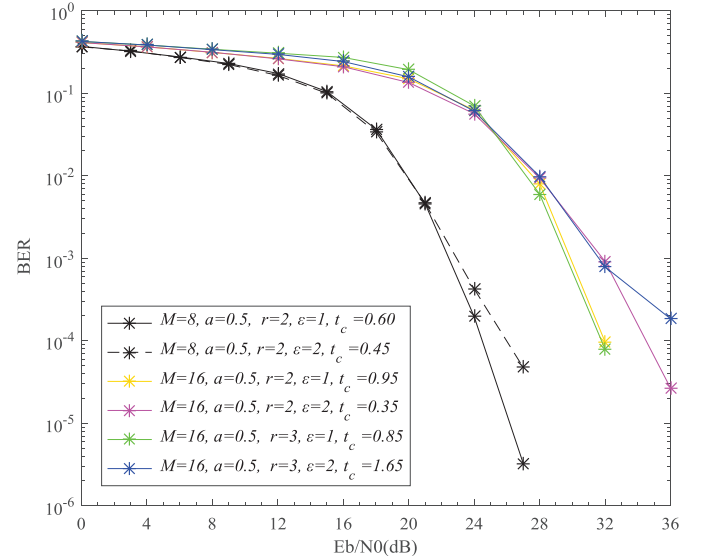}}\\	
	\caption{BER performance of proposed scheme with different $\varepsilon$.}
    \label{fig4}
\end{figure*}

Fig. \ref{Fig4} illustrates the BER performance of the proposed scheme with different values of $\varepsilon$. We can see that the proposed scheme with $\varepsilon=1$ has the best BER performance among all curves. When $\varepsilon \in ({0,1})$, the BER performance will increase as $\varepsilon$ increases. This is because the MED $d_{\min}$ of the superimposed constellation increases with $\varepsilon $ increasing. The $d_{\min}$ reaches its maximum when $\varepsilon=1$, and then it will go down as $\varepsilon>1$ increases.

\begin{figure*}[t]
\centering
	\subfigure[BER performance of optimal $t_x$ over AWGN channel.]{ \label{Fig6}  %
    \includegraphics[width=0.48\textwidth]{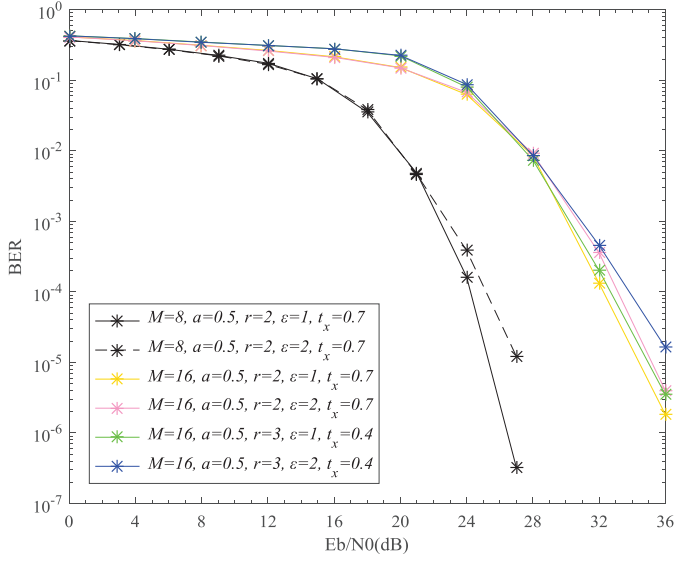}}
    \hspace{2mm}
	\subfigure[BER performance comparison with different $\varepsilon_1$ and $\varepsilon_3$ over AWGN channel when $\lambda=200\% $.]{\label{Fig7}
    \includegraphics[width=0.48\textwidth]{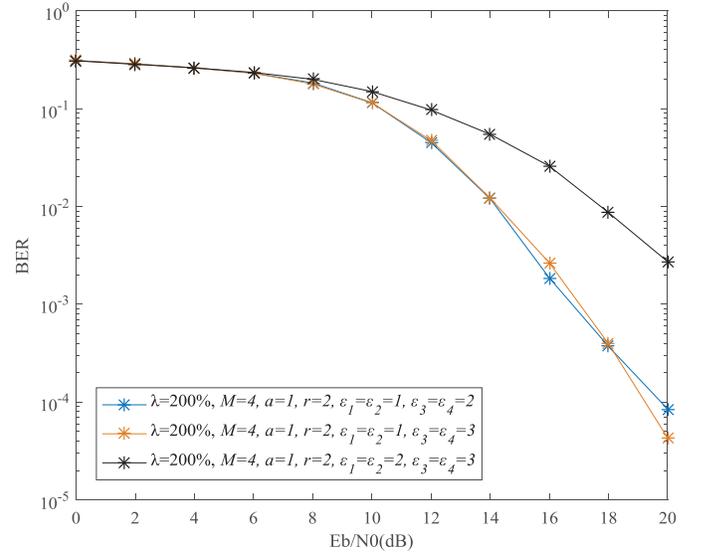}}\\	
	\caption{BER performance of proposed scheme for $\lambda=150\% $ with different $\varepsilon$, and $\lambda=200\% $ with different $\varepsilon_1$ and $\varepsilon_3$.}
    \label{fig5}
\end{figure*}

The BER performance with the optimal $t_c$ and $t_x$ is shown in Fig. \ref{Fig5} and Fig. \ref{Fig6}, respectively. When $r$ is a constant, we can see that different values of $\varepsilon$ will lead to different values of $t_c$, while $t_x$ has the same value for different values of $\varepsilon$. When $M=8$, the BER performance with $t_x$ is better than $t_c$, keeping $\varepsilon$ and $r$ constant. However, the BER performance of $t_c$ is better than $t_x$ with $\varepsilon=1$ and $M=16$. The main reason is that $t_x$ is a more important factor than $t_c$ in the codebook design when $M$ is relatively small, while $t_c$ is a more important factor than $t_x$ in the codebook design when $M$ is relatively large.

Fig. \ref{Fig7} compares the BER performance with different $\varepsilon_1$ and $\varepsilon_3$ when $\lambda=200\% $ (i.e, $K=5$ and $J=10$). It can be seen that the cases for $\varepsilon_1 = \varepsilon_2=1$, $\varepsilon_3= \varepsilon_4 =2$, and $\varepsilon_1 = \varepsilon_2 =1$, $\varepsilon_3 = \varepsilon_4 =3$ have similar BER performance. The BER performance with $\varepsilon_1 = \varepsilon_2 =2$ and $\varepsilon_3 = \varepsilon_4 =3$ is the worst among all three simulation cases. The reason is that the MED of the UDCG-based superimposed constellation will decrease rapidly if both $\varepsilon_1$ and $\varepsilon_3$ take large values.

\subsection{BER Performance of Coded SCMA systems}
In this subsection, we evaluate the BER performance of polar code-based coded SCMA systems with $M=4$, $M=8$ and $M=16$. At the receiver side, a low complexity joint iterative detection and decoding (JIDD) proposed in \cite{8463448} with $5$ iteration times is used for the coded SCMA systems of rate $\frac{1}{2}$. In these simulations, the settings of other parameters are the same as Fig. \ref{Fig1}.

\begin{figure*}[t]
\centering
	\subfigure[BER performance of coded SCMA systems over AWGN channel.]{ \label{fig9a}  %
    \includegraphics[width=0.48\textwidth]{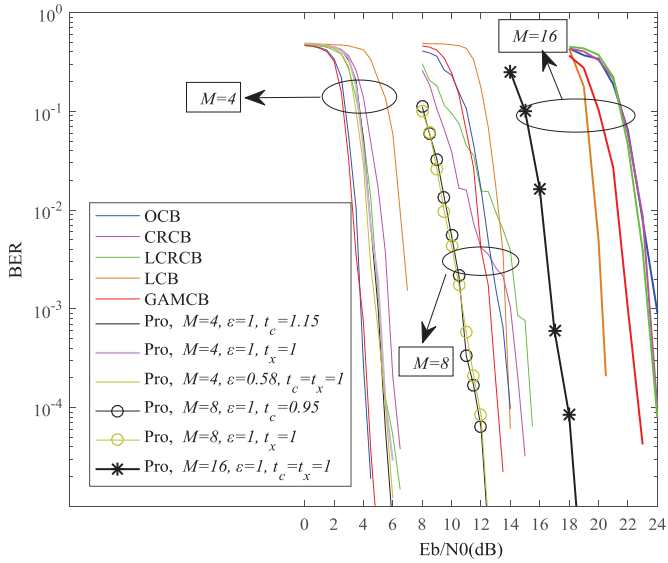}}
    \hspace{2mm}
	\subfigure[BER performance of coded SCMA systems over RF channel.]{\label{fig9b}
    \includegraphics[width=0.48\textwidth]{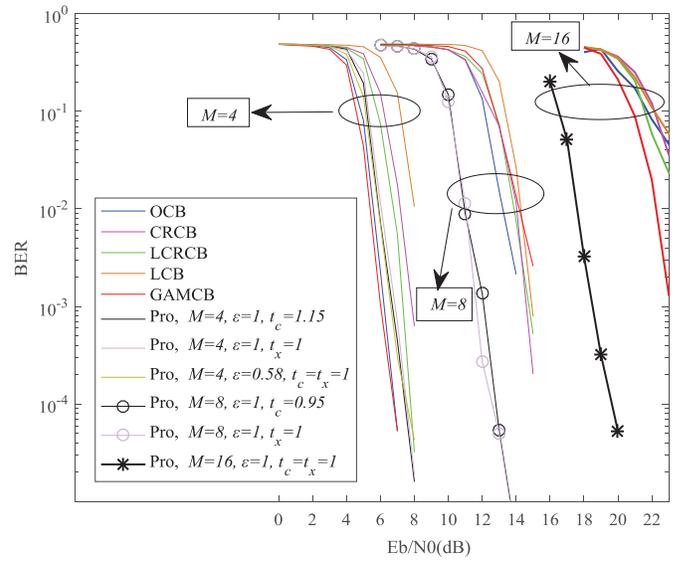}}\\	
	\caption{BER performance of coded SCMA systems with $M=4$, $M=8$ and $M=16$ over AWGN and RF channels.}
    \label{fig6}
\end{figure*}

Fig. \ref{fig6} shows the BER performance of coded SCMA systems with $M=4$, $M=8$ and $M=16$ over AWGN and RF channels. We set $r=1$ in these simulations. From Fig. \ref{fig9a}, we can see that when $M=4$, the BER performance of OCB is worse than that of GAMCB with low Eb/N0, but the best in a high Eb/N0 region. Although the BER performance of the proposed scheme exceeds that of LCB, CRCB and LCRCB, it has 1.5 dB loss compared to OCB scheme. However, the BER performance merit of the proposed codebooks will emerge as $M$ increases. When $M=8$, we see that CRCB scheme has the best BER performance with low Eb/N0, while GAMCB has the best BER performance with high Eb/N0. However, the proposed scheme has nearly 2 dB gain compared to the GAMCB scheme when BER = $10^{-4}$. When $M=16$, the proposed scheme will obtain roughly 3.5 dB gain than LCB scheme, and 5 dB gain than GAMCB scheme when BER = $10^{-3}$. In addition, OCB scheme has the best BER performance when $M=4$, but its BER performance is the worst when $M=16$. Our proposal can keep good BER performance no matter what size of the codebook.

From Fig. \ref{fig9b}, we see that when $M=4$, the BER performance of GAMCB is the best. The proposed scheme has 1 dB loss compared to GAMCB and OCB schemes, but its BER performance exceeds that of LCRCB, CRCB and LCB schemes. Giving $M=8$, the BER performance of OCB is the best, and it can obtain 1.5 dB gain than GAMCB when BER = $10^{-3}$. However, the proposed scheme can achieve extra gain 2 dB higher than OCB design scheme. When $M=16$, GAMCB scheme has the best BER performance. Compared to GAMCB scheme, the proposed scheme can obtain extra 4.5 dB gain. Combining Fig. \ref{fig9a}, we can see that when $M=8$, compared with the existing design schemes, the BER performance of GAMCB is the best over AWGN channel, but is the worst over RF channel. The merit of our proposal is that the proposed scheme can maintain good performance over both AWGN and RF channels.

\begin{figure*}[t]
\centering
	\subfigure[BER performance of the proposed scheme with different $r$, when $M=4$ and $M=8$.]{ \label{fig10a}  %
    \includegraphics[width=0.48\textwidth]{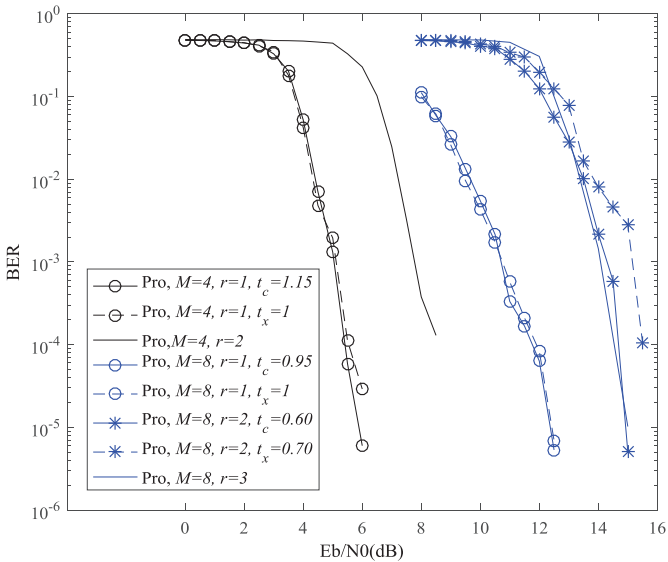}}
    \hspace{2mm}
	\subfigure[BER performance of the proposed scheme with different $r$, when $M=16$.]{\label{fig10b}
    \includegraphics[width=0.48\textwidth]{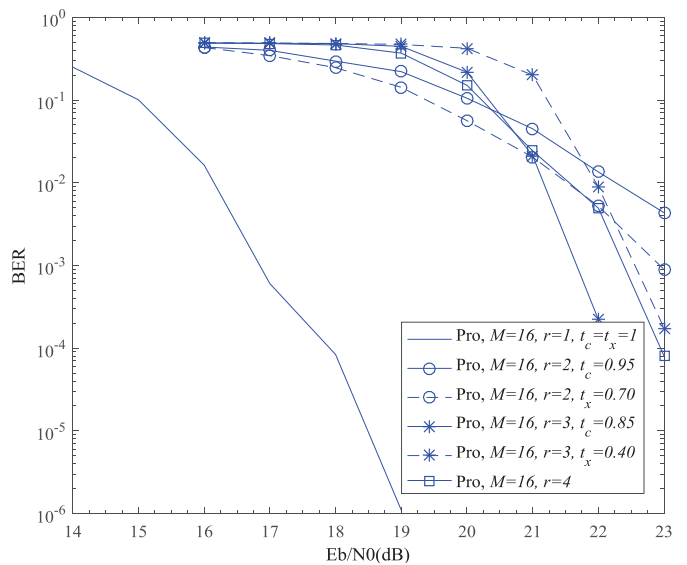}}\\	
	\caption{BER performance of the proposed scheme with different $r$ over AWGN channel.}
    \label{fig7}
\end{figure*}

Fig. \ref{fig7} shows the BER performance of the proposed scheme with different $r$ over AWGN channel. From Fig. \ref{fig10a}, we see that the proposed scheme with $r=1$ has better BER performance than that of the proposed scheme with $r=2$ when $M=4$ and $M=8$. The BER performance of the proposed scheme with $r=1, t_c=1.15$ will outperform that of the proposed scheme with $r=1, t_x=1$ in a high Eb/N0 region. Similar results can be seen from the BER curves of $M=8$, and the difference between the proposed scheme with $t_c$ and $t_x$ becomes large as $r$ increases. When $M=8, r=3$, the proposed scheme has the worst BER performance with low Eb/N0, but its BER will decrease rapidly as Eb/N0 increases, and obtain the same value as the proposed scheme with $r=2, t_c=0.60$. When $M=16$, we can see from Fig. \ref{fig10b} that the BER performance of the proposed scheme with $r=1$ is far superior compared to the proposed scheme with $r>1$. In low Eb/N0 regions, the BER performance of the proposed scheme will be worse with $r$ increasing.

Furthermore, the MED $\left( {d_{\min }^{t_c}}/{d_{\min }^{t_x}} \right)$ of the UDCG-based superimposed constellation is shown in Tab. \ref{tab2}, where ${d_{\min }^{t_c}}$ and ${d_{\min }^{t_x}}$ denote the MED of superimposed constellation with $t_c$ and $t_x$, respectively. From Tab. \ref{tab2}, we see that ${d_{\min }^{t_c}}$ is larger than ${d_{\min }^{t_x}}$ with the same $r$ and $t_c \ne t_x$. With $r$ increasing, the trends of ${d_{\min }^{t_c}}$ and ${d_{\min }^{t_x}}$ are going down. When $r$ is small, large MED of superimposed constellation indicates that large MPD of combination constellation can be obtained by interleaving. Thus, the BER performance of SCMA systems is improved. Combining Tab. \ref{tab1}, we can find that the MED of other design schemes is close to zero with a large $M$. However, the proposed design scheme can still guarantee a large MED. That is why our proposal has better BER performance than that of the existing codebook design schemes, especially for large-size codebooks.

\begin{table*}[t]
\centering
\caption{$d_{\min}^{c,k}$ of different design schemes with $M=4$, $M=8$ and $M=16$.}\label{tab2}
\begin{tabular}{lccc}
\hline
Category & {$M=4 \left( {d_{\min }^{t_c}}/{d_{\min }^{t_x}} \right)$} & {$M=8 \left( {d_{\min }^{t_c}}/{d_{\min }^{t_x}} \right)$} & {$M=16 \left( {d_{\min }^{t_c}/{d_{\min }^{t_x}}} \right)$} \\
\hline
{$r=1$} &0.0752/0.0543 &0.0100/0.0078 & 0.0027/0.0027 \\
{$r=2$} &0.0412        &0.0130/0.0008 & 0.0014/0.0002 \\
{$r=3$} &$-$           &0.0037        & 0.0012/0.0000 \\
{$r=4$} &$-$           &$-$           & 0.0002 \\
\hline
\end{tabular}
\end{table*}

\vspace{0.125in}
\section{Conclusion}
In this paper, we proposed an UDCG and proved that the superimposed constellation of sub-constellations contained in the UDCG is uniquely decodable. Afterwards, we proposed an SCMA codebook design based on the proposed UDCG for downlink SCMA systems, and then obtained the desired BER performance by finding an optimal step size $t$ for both single RN and multiple RNs. The effectiveness of the proposed UDCG-based codebooks was verified over AWGN and RF channels. The following conclusions were obtained from this study. (1) Compared to OCB, CRCB, LCRCB, LCB and GAMCB schemes, the proposed codebook design can obtain the best or almost the best BER performance over both AWGN and RF channels. In uncoded SCMA systems, the BER performance of the proposed scheme with $r=1$ is not good in high SNR regions, but its BER performance is the best in coded SCMA systems. (2) Simulation results show that small-size codebooks with good BER performance do not guarantee a good BER performance of large-size codebooks, such as OCB and GAMCB. However, small-size codebooks with inferior BER performance can achiever a good BER performance of large-size codebooks, such as LCB. Our proposed codebooks can guarantee a good BER performance no matter what size of the codebook. (3) Simulation results disclose that the MED $d_{\min}$ of superimposed constellation is an important KPI for SCMA codebook design.

\vspace{0.125in}
\begin{appendices}
\section{Proof of proposition \ref{propos1}}
\setcounter{equation}{0}
\renewcommand\theequation{A.\arabic{equation}}

The proof of Proposition \ref{propos1} can proceed in two steps as follows.

In the first step, we should show that ${\mathcal S_0}={e^{j\frac{{2\pi }}{{{2^r}}}l}}$ and ${\mathcal S_1}={e^{j\left( {\frac{{2\pi }}{{{2^r}}}l + \frac{\pi }{{{2^{r + \varepsilon}}}}} \right)}}$, ${\mathcal S_0}$ and ${\mathcal S_2}={e^{j\left( {\frac{{2\pi }}{{{2^r}}}l - \frac{\pi }{{{2^{r + \varepsilon}}}}} \right)}}$ form two UDCP groups. In other words, we need to prove that ${\mathcal S_0} = {e^{j\frac{{2\pi }}{{{2^r}}}l}}$ and ${\mathcal S_{1,2}} = {e^{j(\frac{{2\pi }}{{{2^r}}}l \pm \frac{\pi }{{{2^{r + \varepsilon}}}})}}$ satisfy Definition \ref{def3}.

Let ${\alpha _0}, {\alpha _{1,2}} \in A$, ${s_0} = {e^{j\frac{{2\pi }}{{{2^r}}}{l_0}}}$, ${{\tilde s}_0} = {e^{j\frac{{2\pi }}{{{2^r}}}\tilde l_0}}$ in constellation $\mathcal S_0$, and ${s_{1,2}} = {e^{j\frac{{2\pi }}{{{2^r}}}{l_{1,2}}}} {e^{j\theta }}$, ${{\tilde s}_{1,2}} = {e^{j\frac{{2\pi }}{{{2^r}}}{{\tilde l}_{1,2}}}}{e^{j\theta }}$ in constellation $\mathcal S_{1,2}$, where $0 \leqslant {l_0},{\tilde l_0},{l_{1,2}},{\tilde l_{1,2}} \leqslant {2^r} - 1$. Clearly, ${\alpha _0}{s_0} - {\alpha _0}{\tilde s_0}$ can be expressed as
\begin{equation}
{\alpha _0}({s_0} - {{\tilde s}_0}) = 2j{\alpha _0}\sin \bigg[ {\frac{{\pi ({l_0} - \tilde l_0)}}{{{2^r}}}} \bigg] {e^{j\left[ {\frac{{\pi ({l_0} + \tilde l_0)}}{{{2^r}}}} \right]}}.
\label{eq1}
\end{equation}

Similarly, ${\alpha _{1,2}}{\tilde s_{1,2}} - {\alpha _{1,2}}{s_{1,2}}$ can be expressed as
\begin{equation}
{\tilde s_{1,2}} - {s_{1,2}} = 2j{\alpha _{1,2}}\sin \bigg[ {\frac{{\pi (\tilde l_{1,2} - {l_{1,2}})}}{{{2^r}}}} \bigg] {e^{j\left[ {\frac{{\pi ({l_{1,2}} + \tilde l_{1,2})}}
{{{2^r}}} + \theta } \right]}}.
\label{eq2}
\end{equation}

According to (\ref{eq1}) and (\ref{eq2}), ${\alpha _0}{s_0} + {\alpha _{1,2}}{s_{1,2}} = {\alpha _0}{\tilde s_0} + {\alpha _{1,2}}{\tilde s_{1,2}}$ can be rewritten as
\begin{equation}
\begin{aligned}
&{\alpha _0}\sin \bigg[ {\frac{{\pi ({l_0} - \tilde l_0)}}{{{2^r}}}} \bigg]{e^{j\left[ {\frac{{\pi ({l_0} + \tilde {l_0})}}{{{2^r}}}} \right]}} = \\
&~~~~~~~~~{\alpha _{1,2}}\sin \bigg[ {\frac{{\pi (\tilde l_{1,2} - {l_{1,2}})}}{{{2^r}}}} \bigg]{e^{j\left[ {\frac{{\pi ({l_{1,2}} + \tilde l_{1,2})}}{{{2^r}}} + \theta } \right]}}.
\end{aligned}
\label{eq:proof1}
\end{equation}
It can be seen that (\ref{eq:proof1}) holds when ${l_0} = \tilde l_0$ and ${l_{1,2}} = \tilde l_{1,2}$. When ${l_0} \ne \tilde l_0$ and ${l_{1,2}} \ne \tilde l_{1,2}$, Eq. (\ref{eq:proof1}) must satisfy the following conditions
\begin{equation} \label{eq:con1}
\left\{
\begin{aligned}
&\frac{{\pi ({l_0} - \tilde l_0)}}{{{2^r}}} = \frac{{\pi (\tilde l_{1,2} - {l_{1,2}})}}{{{2^r}}} + 2\mu \pi,\\
&\frac{{\pi ({l_0} + \tilde l_0)}}{{{2^r}}} = \frac{{\pi ({l_{1,2}} + \tilde l_{1,2})}}{{{2^r}}} + \theta  + 2\mu'\pi,
\end{aligned}
\right.
\end{equation}
where $\mu$ and $\mu'$ are two integers when $\mu$ is not equal to $0$. According to (\ref{eq:con1}), we have
\begin{subnumcases}
{}
\left[ {({l_0} - \tilde l_0) - (\tilde l_{1,2} - {l_{1,2}})} \right]\bmod {2^r} \equiv 0, \label{con11}\\
{2^{\varepsilon}}\left[ {({l_0} + \tilde l_0) - (\tilde l_{1,2} + {l_{1,2}})} \right]\bmod {2^{r + \varepsilon}} \equiv  \pm 1 \label{con12}.
\end{subnumcases}
From (\ref{eq:proof1}) and (\ref{con11}), it can be seen that $l_0$, $\tilde l_0$, $l_{1,2}$ and $\tilde l_{1,2}$ can be divided by $2^r$. On the other hand, dividing ${2^{\varepsilon}}\left[ {({l_0} + \tilde l_0) - (\tilde l_{1,2} + {l_{1,2}})} \right]$ by $2^{r+\varepsilon}$, we get the result that contradicts to (\ref{con12}). In other words, (\ref{con11}) and (\ref{con12}) cannot hold at the same time. In this case, ${\mathcal S_0}={e^{j\frac{{2\pi }}{{{2^r}}}l}}$ and ${\mathcal S_1}={e^{j\left( {\frac{{2\pi }}{{{2^r}}}l + \frac{\pi }{{{2^{r + \varepsilon}}}}} \right)}}$ can form an UDCP, and ${\mathcal S_0}$ and ${\mathcal S_2}={e^{j\left( {\frac{{2\pi }}{{{2^r}}}l - \frac{\pi }{{{2^{r + \varepsilon}}}}} \right)}}$ can form an UDCP, too.

\vspace{0.15in}
In Step 2, let ${s_1} = {e^{j\left( {\frac{{2\pi }}{{{2^r}}}{l_1} + \frac{\pi }{{{2^{r + \varepsilon}}}}} \right)}}$ and ${{\tilde s}_1} = {e^{j\left( {\frac{{2\pi }}{{{2^r}}}{\tilde l_1} + \frac{\pi }{{{2^{r + \varepsilon}}}}} \right)}}$ be the entries from $\mathcal S_1$, and ${s_2} = {e^{j\left( {\frac{{2\pi }}{{{2^r}}}{l_2} - \frac{\pi }{{{2^{r + \varepsilon}}}}} \right)}}$ and ${{\tilde s}_2} = {e^{j\left( {\frac{{2\pi }}{{{2^r}}}{\tilde l_2} - \frac{\pi }{{{2^{r + \varepsilon}}}}} \right)}}$ be the entries from constellation $\mathcal S_2$, where $0 \leqslant {l_1},{\tilde l_1},{l_2},{\tilde l_2} \leqslant {2^r} - 1$. In what follows, we will show that ${\mathcal S_1}$ and ${\mathcal S_2}$ can form an UDCP, i.e.,
${\alpha _1}{s_1} - {\alpha _1}{\tilde s_1} = {\alpha _2}{\tilde s_2} - {\alpha _2}{s_2}$, where ${\alpha _1}, {\alpha _2} \in A$. According to the proof in the first step, it can be derived that equality ${\alpha _1}{s_1} - {\alpha _1}{\tilde s_1} = {\alpha _2}{\tilde s_2} - {\alpha _2}{s_2}$ holds when ${l_1} = \tilde l_1$ and ${l_2} = \tilde l_2$. When ${l_1} \ne \tilde l_1$ and ${l_2} \ne \tilde l_2$, the condition of ${\alpha _1}{s_1} - {\alpha _1}{\tilde s_1} = {\alpha _2}{\tilde s_2} - {\alpha _2}{s_2}$ can be expressed as
\begin{subnumcases}
{}
\left[ {({l_1} - \tilde {l_1}) - (\tilde {l_2} - {l_2})} \right]\bmod {2^r} \equiv 0, \label{con21}\\
{2^{\varepsilon}}\left[ {({l_1} + \tilde {l_1}) - (\tilde {l_2} + {l_2})} \right]\bmod {2^{r + \varepsilon}} + 2 \equiv 0. \label{con22}
\end{subnumcases}
From (\ref{con21}), we see that ${2^{\varepsilon}}\left[ {({l_1} + \tilde l_1) - (\tilde l_2 + {l_2})} \right]$ can be divided by $2^{r+\varepsilon}$. However, we have $2 < {2^{r + \varepsilon}}$ when $\varepsilon > 0$ and $r$ is a positive integer, which indicates that (\ref{con22}) cannot satisfied. In this case, ${\mathcal S_1}$ and ${\mathcal S_2}$ can form an UDCP.

According to the proof given in Steps 1 and 2, we know that ${\mathcal S_0}$ and ${\mathcal S_1}$, ${\mathcal S_0}$ and ${\mathcal S_2}$, ${\mathcal S_1}$ and ${\mathcal S_2}$, form three UDCPs. In addition, we can derive a conclusion as given in Lemma \ref{lem2}.

\begin{lemma}
If any two constellations in $\{ {{\mathcal S}_n}\} _{n = 0}^N$, ${N \geqslant 1}$, can form an UDCP, then $\{ {{\mathcal S}_n}\} _{n = 0}^N$ form an UDCG.
\label{lem2}
\end{lemma}

The Lemma \ref{lem2} can be proven readily by
\begin{equation}
N \bigg( {\sum\limits_{n = 0}^N {{\alpha _n}{s_n}} } \bigg) = \sum\limits_{n = 0}^N {\sum\limits_{i = n + 1}^N {\left( {{\alpha _n}{s_n} + {\alpha _i}{s_i}} \right)} }.
\end{equation}

According to Lemma \ref{lem2}, we know that the three constellations ${\mathcal S_0}$, ${\mathcal S_1}$, and ${\mathcal S_2}$ form an UDCG. Moreover, we can know that $ C_0$, $ C_1$, and $ C_2$ form an UDCG too by Lemma \ref{lem1}.

This completes the proof of Proposition \ref{propos1}.
\label{appA} \hfill\IEEEQEDclosed

\vspace{0.125in}
\section{Proof of proposition \ref{propos2}}
\setcounter{equation}{0}
\renewcommand\theequation{B.\arabic{equation}}

According to Proposition \ref{propos1}, the condition to support Proposition \ref{propos2} can be expressed as
\begin{equation}
\begin{aligned}
&{\alpha _1}\sin \bigg[ {\frac{{\pi ({l_1} - \tilde l_1)}}{{{2^r}}}} \bigg]{e^{j\left[ {\frac{{\pi ({l_1} + \tilde l_1)}}{{{2^r}}} + {\theta _1}} \right]}} = \\
&~~~~~~~~~~~~~~~{\alpha _2}\sin \bigg[ {\frac{{\pi (\tilde l_2 - l_2)}}{{{2^r}}}} \bigg]{e^{j\left[ {\frac{{\pi ({l_2} + \tilde l_2)}}
{{{2^r}}} + {\theta _2}} \right]}}.
\end{aligned}
\label{p2Con11}
\end{equation}
When ${l_1} = \tilde l_1$ and ${l_2} = \tilde l_2$, Eq. (\ref{p2Con11}) holds obviously. However, the additional condition with ${l_1} \ne \tilde l_1$ and ${l_2} \ne \tilde l_2$ is written as
\begin{equation} \label{eq:p2fend}
\left\{
\begin{aligned}
&\left[ {({l_1} - \tilde {l_1}) - (\tilde {l_2} - {l_2})} \right]\bmod {2^r} \equiv 0, \\
&{2^{{\varepsilon_1} + {\varepsilon_2}}}\left[ {({l_1} + \tilde {l_1}) - (\tilde {l_2} + {l_2})} \right]\bmod {2^{r + {\varepsilon_1} + {\varepsilon_2}}} \pm {2^{{\varepsilon_1}}} \pm {2^{{\varepsilon_2}}} \equiv 0.
\end{aligned}
\right.
\end{equation}
It can be known that Eq. (\ref{eq:p2fend}) can not be held because we have $\left| { \pm {2^{{\varepsilon_1}}} \pm {2^{{\varepsilon_2}}}} \right| < {2^{r + {\varepsilon_1} + {\varepsilon_2}}}$ when $r$ is a positive integer and ${\varepsilon_1} \ne {\varepsilon_2} \geqslant 0$. Therefore, we attain ${l_1} = \tilde l_1$ and ${l_2} = \tilde l_2$ for Proposition \ref{propos2}.

This completes the proof of Proposition \ref{propos2}.
\label{appB}\hfill\IEEEQEDclosed

\vspace{0.125in}
\section{Derivation of $E_c$}
\setcounter{equation}{0}
\renewcommand\theequation{C.\arabic{equation}}

Assume ${l_n} \in \left\{ {0,1, \cdots ,{2^r} - 1} \right\}$ in the $n$-th constellation. It can be known that any constellation point $c_p$ in the superimposed constellation $C_s$ can be expressed as
\begin{equation}
{c_p} = \sum\limits_{n = 0}^{N } {{{\alpha _n}}{e^{j\left( {\frac{{2\pi }}
{{{2^r}}}{l_n}{\text{ + }}{\theta _n}} \right)}}}.  \\
\end{equation}
According to Euler's equation, the energy of $c_p$ can be calculated as
\begin{equation}
{E_p} = \sum\limits_{n = 0}^N {{{\left( {{\alpha _n}} \right)}^2}}  + 2\sum\limits_{n = 0}^N {\sum\limits_{i > n}^N {{\alpha _n}{\alpha _i}} } \cos \bigg[ {\frac{{2\pi }}
{{{2^r}}}\left( {{l_n} - {l_i}} \right) + \left( {{\theta _n} - {\theta _i}} \right)} \bigg].
\end{equation}

Combining the energy of all points together, we get the sum energy $E_c$ as
\begin{equation}
\begin{aligned}
{E_c} =& \sum\limits_{\alpha _0 = a_0}^{a_V} { \cdots \sum\limits_{\alpha _N = a_0}^{a_V} {\left[ {{{\left( 2^r \right)}^{N + 1}}\sum\limits_{n = 0}^N {{\left( {\alpha _n} \right)}^2}  + {{\left( 2^r \right)}^{N - 1}}} \right.} }   \\
& \left. { \times 2\sum\limits_{n = 0}^N {\sum\limits_{i > n}^N {{\alpha _n}{\alpha _i}E} } } \right].
\end{aligned}
\label{eq_energy}
\end{equation}

with $E$ given as
\begin{equation}
\begin{aligned}
E&=\sum\limits_{l_n = 0}^{2^r - 1} {\sum\limits_{l_i = 0}^{2^r - 1} {\cos \left[ {\frac{2\pi }{2^r}\left( {l_n - l_i} \right) + \left( \theta _n - \theta _i \right)} \right]} } \\
&= {2^r}\cos \left( \theta _n - \theta _i \right) + (2^r - 1)\cos \left[ {\frac{2\pi }{2^r} + \left( \theta _n - \theta _i \right)} \right]  \\
&~~~~+(2^r - 1)\cos \left[ { - \frac{2\pi }{2^r} + \left( \theta _n - \theta _i \right)} \right] +  \cdots  + \\
&~~~~\cos \left[ {\frac{2\pi }{2^r}\left( 2^r - 1 \right) + \left( \theta _n - \theta _i \right)} \right] + \\
&~~~~ \cos \left[ { - \frac{2\pi }{2^r}\left( 2^r - 1 \right) + \left(\theta _n - \theta _i \right)} \right] \\
&= \sum\limits_{q = 1}^{2^r} {\left\{ {2q\cos \left[ {\frac{2\pi }
{2^r}\left( 2^r - q \right)} \right]\cos \left( \theta _n - \theta _i \right)} \right\}} \\
&~~~~- {2^r}\cos \left( \theta _n - \theta _i \right).
\label{eq:E}
\end{aligned}
\end{equation}

Moreover, Eq. (\ref{eq_energy}) can be divided into two part, i.e., $E_1$ and $E_2$. The first part $E_1$ can be rewritten as
\begin{equation}
\begin{aligned}
{E_1} &= \sum\limits_{\alpha _0 = a_0}^{a_V} \cdots \sum\limits_{\alpha _N = a_0}^{a_V}\left[ {{{(2^r)}^{N + 1}}\sum\limits_{n = 0}^N {{\left( \alpha _n \right)}^2}} \right]  \\
&= {M^N}{2^r}\left[ {\sum\limits_{n = 0}^N {\sum\limits_{\alpha _n = a_0}^{a_V}{{\left( {\alpha _n} \right)}^2}} } \right].
\end{aligned}
\label{eq:E_1}
\end{equation}

With (\ref{eq:E}) in hand, the second part $E_2$ is rewritten as
\begin{equation}
\begin{aligned}
{E_2} &= \sum\limits_{\alpha _0 = a_0}^{a_V} { \cdots \sum\limits_{\alpha _N = a_0}^{a_V} {\left( {{{\left( 2^r \right)^{N - 1}} \cdot 2}\sum\limits_{n = 0}^N {\sum\limits_{i > n}^N {{\alpha _n}{\alpha _i}E} } } \right)} }   \\
&  = 2{M^{(N - 1)}} \left[ {\sum\limits_{q = 1}^{2^r} {2q\cos \left( {\frac{2\pi q}{2^r}} \right) - 2^r} } \right] \\
&~~~~ \times \left[ {\sum\limits_{n = 0}^N {\sum\limits_{i > n}^N {\sum\limits_{\alpha _n = 0}^{a_V} {\sum\limits_{\alpha _i = 0}^{a_V} {{\alpha _n}{\alpha _i}\cos \left( {\theta _n - \theta _i} \right)} } } } } \right].
\end{aligned}
\label{eq:E_2}
\end{equation}

According to (\ref{eq:E_1}) and (\ref{eq:E_2}), the $E_c$ can be rewritten as
\begin{equation}
\begin{aligned}
E_c &= E_1 + E_2
= {M^N}{2^r}\left[ {\sum\limits_{n = 0}^N {\sum\limits_{\alpha _n = a_0}^{a_V} {{\left( \alpha _n \right)}^2} } } \right] + \\
& ~~~~2{M^{(N - 1)}}\left[ {\sum\limits_{q = 1}^{2^r} {2q\cos \left( {\frac{2\pi q}{2^r}} \right) - 2^r} } \right]  \\
& \times \left[ {\sum\limits_{n = 0}^N {\sum\limits_{i > n}^N {\sum\limits_{\alpha _n = 0}^{a_V} {\sum\limits_{\alpha _i = 0}^{a_V} {{\alpha _n}{\alpha _i}\cos \left( \theta _n - \theta _i \right)} } } } } \right].
\end{aligned}
\label{eq:rw_E}
\end{equation}
Plugging $N=d_f-1$ and $A=[a,a+t, \cdots, a+Vt]^T$ into $(\ref{eq:rw_E})$, we can get
\begin{equation}
\begin{aligned}
E_c &  = {d_f}{M^{(d_f-1)}}{2^r}\left[ {\sum\limits_{v = 0}^V {{\left( a + vt \right)}^2} } \right] + \\
&~~~~ 2{M^{(d_f - 2)}} \left[ {\sum\limits_{q = 1}^{2^r} {2q\cos \left( {\frac{2\pi q}{2^r}} \right) - 2^r } } \right]  \\
&  \times {\left[ {\sum\limits_{v = 0}^V {\left( a + vt \right)} } \right]^2}\left[ {\sum\limits_{n = 0}^{d_f-1} {\sum\limits_{i > n}^{d_f-1} {\cos \left( {\theta _n - \theta _i} \right)} } } \right].
\end{aligned}
\end{equation}

This completes the derivation of $E_c$.
\label{appC}\hfill\IEEEQEDclosed

\vspace{0.125in}
\section{Derivation of two special cases in $d_c$}
\setcounter{equation}{0}
\renewcommand\theequation{D.\arabic{equation}}

Assume that another constellation point $c_p'\ (p' \ne p)$ in superimposed constellation $C_s$ is expressed as
\begin{equation}
{c_p'} = \sum\limits_{n = 0}^{N} {\left( {{{\alpha _n}}{\text{ + }}\Delta _a^n} \right){e^{j\left[ {\frac{{2\pi }}{{{2^r}}}\left( {{l_n} + \Delta _\theta ^n} \right) + {\theta _n}} \right]}}},
\end{equation}
where $\Delta _a^n$ and $\Delta _\theta ^n$ denote the amplitude and angle transform operations for the $n$-th constellation, respectively. If $\Delta _a^n=0$ and $\Delta _\theta ^n=0$, we get $c_p'$=$c_p$. Therefore, the square distance ${\left( {{d_c}} \right)^2}$ between $c_p$ and $c_p'$ can be expressed as
\begin{equation}
\begin{aligned}
{\left( d_c \right)^2} &= \left\{ {\sum\limits_{n = 0}^N {\left[ {\left( {\alpha _n + \Delta _a^n} \right)\cos \left( {\frac{2\pi }{2^r} {l_n + \frac{2\pi }{2^r}\Delta _\theta ^n}  + \theta _n} \right)} \right.} } \right. \\
& - {\left. {\left. { \alpha _n \cos \left( {\frac{2\pi }{2^r}{l_n}}+ \theta _n \right)} \right]} \right\}^2}  \\
&  + \left\{ {\sum\limits_{n = 0}^N {\left[ {\left( {\alpha _n + \Delta _a^n} \right)\sin \left( {\frac{2\pi }{2^r} {l_n + \frac{2\pi }{2^r}\Delta _\theta ^n} + \theta _n} \right)} \right.} } \right. \\
& - {\left. {\left. { \alpha _n \sin \left( {\frac{2\pi }{2^r}{l_n}}+ \theta _n \right)} \right]} \right\}^2},
\end{aligned}
\label{eq:ed}
\end{equation}
where ${\alpha _n} \in A$, ${\alpha _n}+\Delta _a^n \in A$, $l_n \in ( {0,{2^r} - 1} )$ and ${l_n}+\Delta _\theta ^n \in( {0,{2^r} - 1} )$. Furthermore, we obtain $\Delta _a^n \in \left[ { - vt, - \left( {v - 1} \right)t, \cdots ,\left( {V - v} \right)t} \right]$ when ${\alpha _{{n}}}=a+vt$. Similarly, we can derive $\Delta _\theta ^n \in ( { - {l_n},1 - {l_n}, \cdots ,{2^r} - {l_n} - 1})$. For analytical convenience, let
\begin{equation}
\begin{aligned}
g_n^c &= \left( \alpha _n + \Delta _a^n \right)\cos \left( {\frac{2\pi }{2^r}\left( {l_n + \Delta _\theta ^n} \right) + \theta _n} \right) \\
&~~~~ - {\alpha _n}\cos \left( {\frac{2\pi }{2^r}{l_n} + \theta _n} \right) \\
&  =  - 2{\alpha _n}\sin \left( {\frac{2\pi }{2^r}{l_n} + \frac{\pi }
{2^r}\Delta _\theta ^n + \theta _n} \right)\sin \left( {\frac{\pi }
{2^r}\Delta _\theta ^n} \right) \\
&~~~~ + \Delta _a^n\cos \left[ {\frac{2\pi }{2^r}\left( {l_n + \Delta _\theta ^n} \right) + \theta _n} \right],
\end{aligned}
\label{eq:cos}
\end{equation}
\begin{equation}
\begin{aligned}
g_n^s & = \left( \alpha _n + \Delta _a^n \right)\sin \left( {\frac{2\pi }
{2^r}\left( l_n + \Delta _\theta ^n \right) + \theta _n} \right) \\
&~~~~ - {\alpha _n}\sin \left( {\frac{2\pi }{2^r}{l_n} + \theta _n} \right) \\
&=  2{\alpha _n}\cos \left( {\frac{2\pi }{2^r}{l_n} + \frac{\pi }{2^r}\Delta _\theta ^n + \theta _n} \right)\sin \left( {\frac{\pi }{2^r}\Delta _\theta ^n} \right) \\
&~~~~ + \Delta _a^n\sin \left[ {\frac{2\pi }{2^r}\left( l_n + \Delta _\theta ^n \right) + \theta _n} \right].
\end{aligned}
\label{eq:sin}
\end{equation}
In this case, (\ref{eq:ed}) can be simplified as
\begin{equation}
\begin{aligned}
{\left( d_c \right)^2} &= {\left( {\sum\limits_{n = 0}^N {g_n^c} } \right)^2} + {\left( {\sum\limits_{n = 0}^N {g_n^s} } \right)^2} \\
& = \sum\limits_{n = 0}^N {\left[ {{{\left( g_n^c \right)}^2} + {{\left( g_n^s \right)}^2}} \right]}  +  \sum\limits_{n = 0}^{N - 1} {\sum\limits_{i > n}^N {2\left( g_n^c g_i^c + g_n^s g_i^s \right)} }.
\end{aligned}
\label{eq:dFinally}
\end{equation}
From (\ref{eq:dFinally}), two special cases can be derived as follows.
\begin{enumerate}
\item
When $\Delta _a^n=0, \forall n$ and $2^r \leqslant M$, (\ref{eq:dFinally}) can be written as
\begin{equation}
\begin{aligned}
{\left( {d_1^c} \right)^2} &= \sum\limits_{n = 0}^N {\bigg[ {4{{\left( \alpha _n \right)}^2}{{\left( {\sin \frac{\pi }{2^r}\Delta _\theta ^n} \right)}^2}} \bigg]} \\
& + \sum\limits_{n = 0}^{N - 1} {\sum\limits_{i > n}^N {\bigg\{ {8{\alpha _n}{\alpha _i}\sin \left( {\frac{\pi }{2^r}\Delta _\theta ^n} \right)\sin \left( {\frac{\pi }
{2^r}\Delta _\theta ^i} \right)} } }   \\
&  {\cos \bigg[ {\frac{2\pi }{2^r}\left( {l_n - l_i} \right) + \frac{\pi }{2^r}\left( \Delta _\theta ^n - \Delta _\theta ^i \right) + {\theta _n} - {\theta _i}} \bigg]} \bigg\}.
\end{aligned}
\label{eq:speCase1}
\end{equation}

In (\ref{eq:speCase1}), let us consider that only two constellation angles will change, i.e., $\Delta _\theta ^{n} \ne 0$ and $\Delta _\theta ^{i} \ne 0$, where ${n} \ne {i}$, for the $n$-th and the $i$-th constellations, respectively. Let $\Delta _\theta ^{n}$ and $\Delta _\theta ^{i}$ be the opposite operations, i.e., $\Delta _\theta ^{n}+\Delta _\theta ^{i}=0$. In particular, when $\Delta _\theta ^{n}=1$, we can imagine that ${d_{1}^c}=0$ if $\frac{2\pi }{2^r}\left( l_n - l_i \right) + {\theta _n} - {\theta _i}= \frac{2\pi }{2^r}$. Assume ${\theta _{1,\min }^c} = \min \left[ {\left| {\frac{2\pi }{2^r}\left( l_n - l_i - 1 \right) + {\theta _n} - {\theta _i}} \right|} \right]$ and $\Delta _\theta ^{n}=1$. Thus, the minimum value of (\ref{eq:speCase1}) can be calculated as
\begin{equation}
{\left( {d_{1,\min }^c} \right)^2} = 8{a^2}{\left( {\sin \frac{\pi }
{2^r}} \right)^2}\left( {1 - \cos \theta _{1,\min }^c} \right).
\label{eq:case1Finally}
\end{equation}
\item
When $\Delta _\theta ^n=0, \forall n$ and $2^r<M$, (\ref{eq:dFinally}) can be expressed as
\begin{equation}
\begin{aligned}
{\left( d_c \right)^2} &= \sum\limits_{n = 0}^N {{\left( \Delta _a^n \right)}^2}  + \sum\limits_{n = 0}^{N - 1} {\sum\limits_{i > n}^N {\left\{2\Delta _a^n\Delta _a^i \right.} }   \\
&~~~~ \left. { \times \cos \left[ {\frac{2\pi }{2^r}\left( l_n - l_i \right) + {\theta _n} - {\theta _i}} \right]} \right\}.
\end{aligned}
\label{eq:dcase2}
\end{equation}
Moreover, let us consider that two constellations $\Delta _a^{n} \ne 0$ and $\Delta _a^{i}\ne 0$ in (\ref{eq:dcase2}) for the $n$-th and the $i$-th constellations, respectively, where $\Delta _a ^{n}$ and $\Delta _a ^{i}$ denote the opposite operations, i.e., $\Delta _a ^{n}+\Delta _a ^{i}=0$. In this case, (\ref{eq:dcase2}) can be rewritten as
\begin{equation}
\begin{aligned}
{\left( d_2^c \right)^2} & = {\left( \Delta _a^n \right)^2} + {\left( \Delta _a^i \right)^2} + 2\Delta _a^n\Delta _a^i \\
&\times \cos \bigg[ {\frac{2\pi }{2^r}\left( l_n - l_i \right) + {\theta _n} - {\theta _i}} \bigg],
\end{aligned}
\label{eq:twocons}
\end{equation}
which has a minimum value when $\Delta _a^{n}=t$ and $\left| {\frac{2\pi }{2^r}\left( l_n - l_i \right) + \theta _n - \theta _i} \right|$ are minimized. Let ${\theta _{2,\min}^c}= \min\left[ {\left| {\frac{2\pi }{2^r}\left( l_n - l_i \right) + {\theta _n} - {\theta _i}} \right|} \right]$. The minimum value of (\ref{eq:twocons}) can be calculated as
\begin{equation}
\left( d_{2,\min}^c \right)^2 = 2{t^2}\left( {1 - \cos {\theta _{2,\min }^c}} \right).
\label{eq:case2Finally}
\end{equation}
\end{enumerate}

In summary, (\ref{eq:case1Finally}) and (\ref{eq:case2Finally}) give two square distances derived from (\ref{eq:dFinally}), where (\ref{eq:case1Finally}) denotes the MED with $\Delta _\theta^n$ and (\ref{eq:case2Finally}) denotes the MED with $\Delta _a^n$.

This completes the derivation of two special cases in $d_c$.
\label{appD}\hfill\IEEEQEDclosed
\end{appendices}

\vspace{0.125in}
\begin{bibliographystyle}{IEEEtran}
\begin{bibliography}{IEEEabrv,bibtex}
\end{bibliography}
\end{bibliographystyle}

\vfill

\end{document}